\documentclass[manuscript,screen]{acmart}

\AtBeginDocument{%
  }

\setcopyright{acmlicensed}
\copyrightyear{2025}
\acmYear{2025}
\acmDOI{XXXXXXX.XXXXXXX}

\acmISBN{978-1-4503-XXXX-X/2018/06}

\usepackage[english]{babel}
\usepackage{graphicx}
\usepackage{multicol}
\usepackage{footmisc}
\usepackage{multirow}
\usepackage{booktabs}
\usepackage{lipsum}
\usepackage{enumerate}
\usepackage{rotating}
\usepackage{tikz}
\usepackage{bbm}
\usepackage{placeins}

\usepackage{amssymb}
\usepackage{tabularx}
\usepackage{hyperref}
\usepackage{algorithm}
\usepackage{algpseudocode}
\usepackage{makecell}

\usepackage{soul}  

\begin{document}

\title{Self-Supervised Multi-Stage Domain Unlearning for White-Matter Lesion Segmentation}

\author{Domen Prelo\v{z}nik}
\email{domen.preloznik@fe.uni-lj.si}
\orcid{0009-0006-5829-1578}
\author{\v{Z}iga \v{S}piclin}
\authornote{Corresponding author.}
\email{ziga.spiclin@fe.uni-lj.si}
\orcid{0000-0001-8300-0417}
\affiliation{%
  \institution{University of Ljubljana, Faculty of Electrical Engineering}
  \city{Ljubljana}
  \state{Slovenia}
  \country{Slovenia}
}

\renewcommand{\shortauthors}{Prelo\v{z}nik and \v{S}piclin}

\begin{abstract}
Inter-scanner variability of magnetic resonance imaging  has an adverse impact on the diagnostic and prognostic quality of the scans and necessitates the development of models robust to domain shift inflicted by the unseen scanner data. Review of recent advances in domain adaptation showed that efficacy of strategies involving modifications or constraints on the latent space appears
to be contingent upon the level and/or depth of supervision during model training. %
In this paper, we therefore propose an unsupervised domain adaptation technique based on self-supervised multi-stage unlearning (SSMSU). Building upon the state-of-the-art segmentation framework nnU-Net, we employ deep supervision at deep encoder stages using domain classifier unlearning, applied sequentially across the deep stages to suppress domain-related latent features. Following self-configurable approach of the nnU-Net, the auxiliary feedback loop implements a self-supervised backpropagation schedule for the unlearning process, since continuous unlearning was found to have a detrimental effect on the main segmentation task.
Experiments were carried out on four public datasets for benchmarking white-matter lesion segmentation methods. Five benchmark models and/or strategies, covering passive to active unsupervised domain adaptation, were tested. 
In comparison, the SSMSU demonstrated the advantage of unlearning by enhancing
lesion sensitivity and limiting false detections, which resulted in higher overall segmentation quality in terms of segmentation overlap and relative lesion volume error. The proposed model inputs only the FLAIR modality, which simplifies preprocessing pipelines, eliminates the need for inter-modality registration errors and harmonization, which can introduce variability.  Source code is available on \url{https://github.com/Pubec/nnunetv2-unlearning}.
\end{abstract}

\ccsdesc[500]{Computing methodologies~Image segmentation}
\ccsdesc[500]{Applied computing~Health care information systems}

\keywords{Scanner Variability, Image Segmentation, Unsupervised Domain Adaptation, Self-Supervised Unlearning, Comparative Evaluation, Open-Source and Reproducible}


\maketitle

\section{Introduction}

\begin{figure}[!b]
  \centering
  \includegraphics[width=0.75\textwidth]{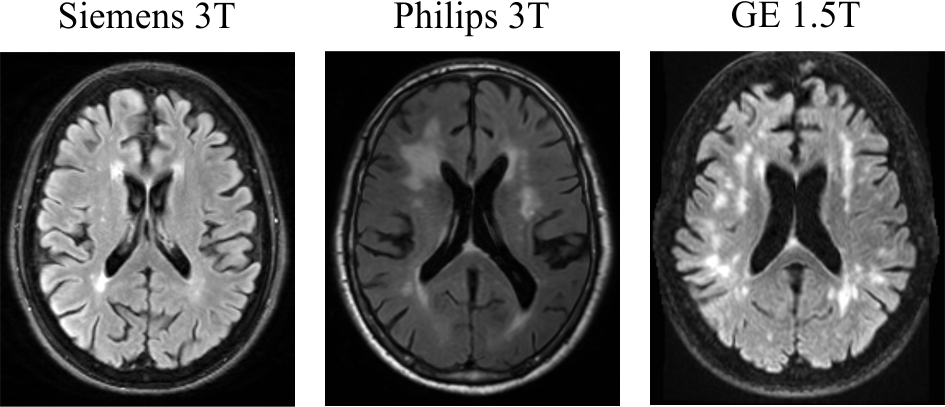}
  \caption{Comparison of MRI scans between different domains.} \label{figure:dataset-comparison}
\end{figure}

Proliferation of brain Magnetic Resonance Imaging (MRI) in conjunction with deep learning (DL) image analysis enables state-of-the-art white-matter lesion (WML) segmentation~\citep{amrita_sota_wml_seg_2020,eshaghi_identifying_2021, mcginley_diagnosis_2021}. Segmentation should be accurate and robust as the total number and volume of lesions quantified from MRI were established as biomarkers for diagnosis and response to therapy assessment~\citep{popescu_brain_2013}, and future Multiple Sclerosis (MS) disease progression~\citep{oship_assessment_2022}. 
However, inter-scanner variabilities in intensity contrasts and noise distributions (Figure \ref{figure:dataset-comparison}) persist as most detrimental factors impacting the diagnostic and prognostic quality of the MRI scans~\citep{sastregarriga-2020}. They adversely affect a segmentation model's ability to generalize to new domains (i.e., new scanners), whereas obtaining high quality expert lesion annotations on the new domain data for model fine-tuning is often prohibitively time consuming and impractical. This  
necessitates the development of broadly applicable segmentation models that are not restricted to the source domain, on which they were trained.

The difficulty of consistently analyzing data from unseen domain is increasingly highlighted in the DL literature. For example, \cite{alzubaidi_review_2021} and the Google Research Team \citep{damour_underspecification_2022} refer to \textit{underspecification}, where they explain poor behavior of DL systems when deployed to real-world domains, regardless of the area of application. 
When evaluating WML segmentation, the same phenomenon is apparent in public WML segmentation challenge results. In MICCAI 2016 MS Lesion Segmentation Challenge (MSSEG) results \citep{commowick_objective_2018} the segmentation performances of all tested methods generalized well on test set of the source scanner data, on which they were trained, but dropped substantially on the unseen test set (\textit{Center 3}). 
Similar observations hold for the MICCAI 2021 MS New Lesions Segmentation Challenge (MSSEG-2) results \citep{commowick_hal_03358968}. 
Even the top ranking team's method exhibited a drop in DSC score from 0.52 (\textit{seen}) to 0.26 (\textit{unseen}). The domain shift therefore needs to be addressed accordingly.

We hypothesize that poor DL model performance generalization across domains and domain overfitting mainly arise from the inability to discriminate between the task-related and domain-related features, and their respective enhancement instead of suppression, but also other factors, such as inadequate or insufficient data augmentation, use of MRI preprocessing and/or harmonization, data imbalance, to mention a few. 
In order to address the challenge of unsupervised domain adaptation (UDA) in DL models we propose a novel self-supervised multi-stage domain unlearning strategy and demonstrate comparatively enhanced performance for the task of WML segmentation.

\section{Related Work and Contributions}\label{section:related-work}

Domain adaptation is a specific type of transfer learning, in which domain features -- a specific feature subspace defined by source dataset -- and task features defined by labels and goals, are disentangled so as to enable knowledge transfer, e.g., emphasizing task features and disregarding/suppressing domain-identifying features. This adaptation allows models to adequately apply prior knowledge to previously unseen target domain~\citep{guan2022survey}.

Data and model diversification is a prospective form of UDA to aid model generalizability. Simple techniques involve aggressive data augmentation and generation of synthetical data during training~\citep{8995481, basaran2024seghedsegmentationheterogeneousdata,cackowski_imunity_2023}, provided that domain-identifying variations can be modeled reasonably well. In LST-AI \citep{wiltgen_lst-ai_2024} that is based on an ensemble of three 3D U-Nets~\citep{ronneberger_u-net_2015,yang_densely_2021}, each trained with deep supervision at one or two layers and various combinations of losses, promoting model diversity showed good generalization performance over state-of-the-art models on unseen datasets.

Data harmonization based UDA techniques aim to reduce the variability caused by differences in acquisition protocols, scanners, or sites.  
Common prospective form is MRI preprocessing~\citep{akkus_deep_2017}. It may include registration to standard brain atlas space, non-brain tissue removal, bias field correction, denoising and intensity normalization. The preprocessing is standardized in many public segmentation challenges, which usually provide preprocessed scans alongside raw ones; e.g. \citep{Kuijf2022_miccai2017wmh,commonwick2021miccai2016,commowick_hal_03358968}. The downside is that with intensity manipulation such as bias correction or denoising one may inadvertently remove task-relevant features, potentially degrading downstream segmentation model effectiveness.

Voxel-level harmonization methods such as Ravel~\citep{fortin_removing_2016}, ComBat and its extensions~\citep{fortin_harmonization_2017,pomponio_harmonization_2020,beer_longitudinal_2020} have been successfully applied to retrospectively evaluate and compensate harmonization effects in diffusion MRI and longitudinal structural imaging, but are applied retrospectively. Image-level harmonization, such as MURD~\cite{liu_learning_2024}, seeks to translate scanner-specific appearance while preserving anatomy, but exposes the risk of altering or hallucinating anatomically relevant details that is particularly problematic for pathology-sensitive tasks like WML segmentation. In general, these techniques require recalibration each time new scanners or acquisition sites are added to the dataset and rely on demographic subject information, which is not always available.

Feature-level methods, such as MSCDA~\cite{kuang_mscda_2023}, align representations via contrastive objectives, but rely on carefully designed positive/negative pairs and often assume access to all domains at inference, limiting their clinical applicability. Domain-randomized synthetic training, exemplified by SynthSeg~\cite{billot_synthseg_2023}, offers scalability yet may fail to capture realistic pathological variability due to reliance on fully synthetic images. 

Generative methods have also been proposed to address domain shift in MRI. ImUnity~\citep{cackowski_imunity_2023} is a self-supervised VAE-GAN–based harmonization approach that removes scanner-specific biases and preserves clinical features, but is currently limited to T1-weighted scans and may introduce task-irrelevant features due to inherent choice of target domain in  decoder. SAMSEG~\citep{cerri_contrast-adaptive_2021} uses a forward probabilistic model with latent variables to constrain WML shape and location for automated segmentation. However, generative and harmonization methods share key limitations: \textbf{(i)} reduced robustness in cases with marked biological or pathological variability (e.g., severe atrophy or extensive WMLs), and \textbf{(ii)} they do not consistently improve downstream segmentation performance.

Recent implicit, representation-free techniques like domain unlearning involve training a domain classification network, based on encoder and/or decoder outputs of the main task network, and its unlearning via domain classifier confusion loss to nullify encoding of domain-related information~\citep{dinsdale2021unlearning, guan_multi-site_2021}. This is analog to the confusion module at the bottleneck layer in ImUnity~\citep{cackowski_imunity_2023}, a self-supervised Variational AutoEncoder developed for T1-weighted MRI scan normalization. Another approach is by fusing latent space with respect to domain-identifying features, via latent embeddings, while contrasting them for the main task~\citep{wolleb2022learn2ignore}. Similarly, a disentanglement of the anatomical and domain feature representations was achieved by penalizing feature map interdependence via mutual information minimization~\citep{bi2024misegnet}. According to these advances, we hypothesize that the efficacy of strategies involving modifications or constraints on the latent space depends on the level and/or depth of model supervision during training. 

The identified limitations of aforementioned UDA strategies motivate our focus on representation-level unlearning, which avoids generative translation assumptions and remains compatible with heterogeneous multi-scanner FLAIR datasets in real-world scenario.

Efficacy of the segmentation advances should be determined through a comparative analysis with state-of-the-art segmentation frameworks such as nnU-Net~\citep{isensee2021nnunet}, Transformer networks~\citep{hatamizadeh2022swin}, or their ensembles~\citep{alshehri_ischemic_2024}, which is not always the case. As pointed out in \cite{isensee2024nnunetrevisitedrigorousvalidation} and \cite{confidenceintervalsuncoveredready}, the use of inadequate baselines may leave a gap in understanding of the relative effectiveness of the novel approaches. While based on simple U-net architecture, but with aggressive data augmentation, the nnU-Net proved surprisingly effective in addressing domain adaptation in an unsupervised manner. We aim to address the same goal, i.e. with model training performed on \textit{seen}, while testing on \textit{unseen} domain, in the context of WML segmentation in MRI scans of different scanner vendors and sites.

\subsection{Contributions}
In this paper, we enhance the nn-Unet segmentation framework with multi-stage self-configuring deep supervision at each convolutional-block of the encoder using domain classifier unlearning, applied across multiple layers, so as to suppress the propagation of domain-related latent features at each encoder layer that could potentially hamper model generalization ability to the unseen domain. 

This paper makes five key contributions:
\begin{itemize}
    \item[(1)] We propose a \textit{multi-stage self-supervised unlearning} strategy that suppresses scanner-specific information without requiring target-domain data (unsupervised/zero-shot DA);
    \item[(2)] We integrate this strategy into nnU-Net~\citep{isensee2021nnunet}, providing a general and open-source solution for domain-affected segmentation tasks;
    \item[(3)] We perform a reproducible evaluation on four public multi-scanner FLAIR MRI datasets for WML segmentation, covering both seen and unseen domains;
    \item[(4)] We show that common \textit{intensity-based preprocessing} (denoising, bias correction) can hinder segmentation and reduce the effectiveness of UDA strategies;
    \item[(5)] We demonstrate that \textit{minimal preprocessing} enables stronger benefits from the proposed unlearning method and data augmentation, yielding consistent cross-domain improvements.
\end{itemize}

\section{Materials and Methods}

\subsection{Datasets}

Four public MRI datasets containing cases with WMLs were employed in this study, namely the White-Matter Hyperintensities (WMH) 2017 Segmentation Challenge \citep{Kuijf2022_miccai2017wmh}, the MSSEG  \citep{commonwick2021miccai2016}, MS Ljubljana dataset (MSLJ)~\citep{lesjak_novel_2018} and ISBI 2015 training dataset~\citep{Carass2017}. 
Details about datasets are summarized in Table \ref{table:wmh-data-description}, with reported train/test data splits as used later in experimental evaluation. 

\subsection{Preprocessing}
All FLAIR scans were preprocessed from raw acquisitions in the same manner: \textbf{(i)} linearly registered to the T1-weighted MNI 2009c brain atlas space~\citep{fonov2009mniatlas}, \textbf{(ii)} intensity range linearly rescaled to [0-1] interval, and \textbf{(iii)} cropped to size $192\times 224\times 192$, containing the whole brain. Such preprocessing pipeline standardized the scan dimensions and intensity range, but intentionally did not modify the tissue contrasts, bias field, and noise distributions.
To assess the impact of intensity preprocessing (itself a form of UDA) on cross-domain performance of the WML segmentation models, we additionally evaluated an extended preprocessing pipeline, that included the aforementioned three steps and subsequently applied: \textbf{(iv)} N4 bias correction~\citep{tustison_n4itk_2010}, \textbf{(v)} adaptive non-local means denoising~\citep{manjon_adaptive_2010}, and \textbf{(vi)} non-brain tissue removal. %
Computations were performed using the AntsPy package\footnote{AntsPy: \url{https://github.com/ANTsX/ANTsPy}} \citep{ants}. 

\begin{table}[!bht]
  \caption{Dataset information and train / test splits as used in experimental evaluation.}
  \label{table:wmh-data-description}
  \centering
  \begin{tabular}{|l|l|c|r|r|r|}
    \bottomrule[0.5pt]

    \textbf{Source / Subset} & \textbf{Vendor} & \textbf{FS} & \textbf{AM} & \textbf{Train} & \textbf{Test} \\
    
    \toprule[0.5pt]
    \bottomrule[0.5pt]
    
    \multirow{2}{*}{WMH / Seen}
      & Philips & 3T & 2D & 20 & 30~ \\ 
      & Siemens & 3T & 2D & 20 & 30~ \\
    \toprule[0.5pt]
    \bottomrule[0.5pt]
    \multirow{3}{*}{WMH / Unseen}
      & Philips & 3T & 3D &  & 10~ \\ 
      & GE & 3T & 3D &  & 30~ \\ 
      & GE & 1.5T  & 3D &  & 10~ \\ \hline\hline 
    \multirowcell{4}{MSSEG /\\Unseen}
    & Philips & 3T   & 3D &  & 15~ \\ 
    & Siemens & 3T   & 3D &  & 15~ \\ 
    & Siemens & 1.5T & 3D &  & 14$\dagger$ \\ 
    & GE      & 3T   & 3D &  & 8~ \\ \hline\hline
    \multirow{1}{*}{MSLJ / Unseen} 
    & Siemens & 3T   & 3D &  & 30~ \\ \hline\hline
    \multirow{1}{*}{ISBI / Unseen} 
    & Philips      & 3T   & 2D &  & 21~ \\ \hline
    \multicolumn{6}{l}{\scriptsize FS--Field Strength; AM--Acquisition Mode;
    $\dagger$Originally 15, but one case} \\[-3pt]
    \multicolumn{6}{l}{\scriptsize (\texttt{msseg-test-center07-08}) was removed due to incorrect reference mask.} \\[-3pt]
  \end{tabular}
\end{table}

\subsection{nnU-Net baseline}\label{subsection:nnunet-baseline}

\begin{figure*}[ht]
  \centering
  \includegraphics[width=1.0\textwidth]{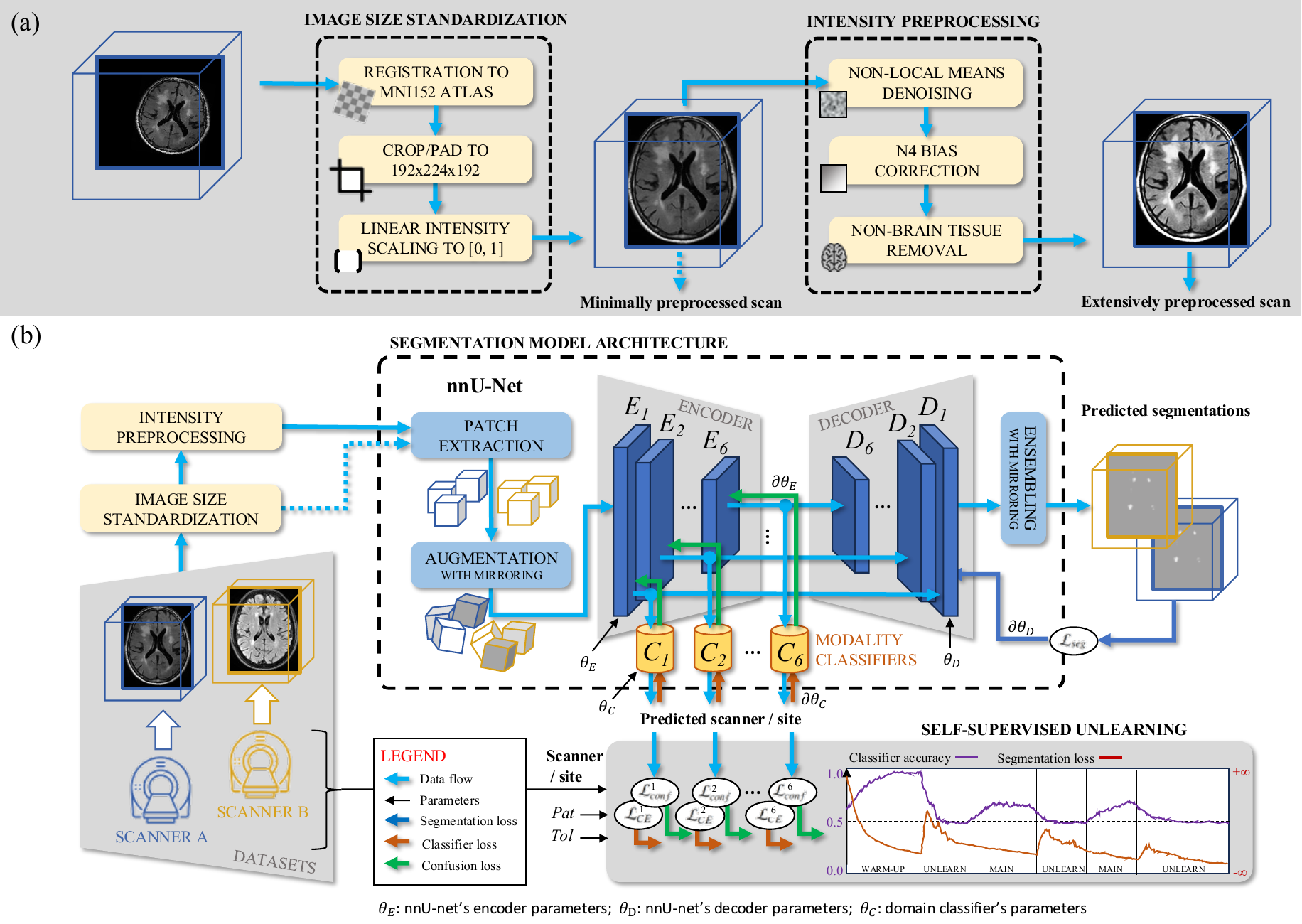}
  \caption{(a) MRI preprocessing and (b) self-supervised multi-stage unlearning with nnU-Net.} \label{figure:architecture}
\end{figure*}

Our baseline model was the default nnU-Net self-confi-guring framework\footnote{nnU-Net: \url{https://github.com/MIC-DKFZ/nnUNet}}.
The \textit{3D-fullres} model is depicted in Figure ~\ref{figure:architecture}, with encoder consisting of six stages (denoted with \textit{E1--6}), each consisting of two consecutive Conv blocks of 3$\times$3$\times$3 convolutional layer, instance normalization (IN) and leaky rectified linear unit (lReLU) activation. In top stage the first block had stride 1 in the convolutional layer, while respective stride was 2 in other stages to achieve downsampling.
The decoder consisted of six stages \textit{D1--6}, inputting skip connections between respective encoder block \textit{E1--6} at same resolution. Each decoder block consisted of 2$\times$2$\times$2 transpose convolution with stride 2, IN and lReLU activation followed by a Conv block. 
For training segmentation models, a Dice Similarity Coefficient (DSC) based loss \citep{milletari_v-net_2016_diceloss} is a common (non)overlap measure defined as:
\begin{equation}
  \mathcal{L}_{\text{DSC}} = 1 - \frac{2 \times \text{TP}}{2 \times \text{TP} + \text{FP} + \text{FN}}\,,
  \label{eq:dsc}
\end{equation}
\noindent
where TP, FP, and FN denote the true and false positive, and false negative voxel labels, respectively.
Similarly, the Cross Entropy (CE) loss  \citep{Kline2005RevisitingSA_celoss} measures pixel-wise classification performance:
\begin{equation}
    \mathcal{L}_{CE} =-\sum_{n=1}^{N} t_n \log y_n\,,
  \label{eq:cross_entropy}
\end{equation}
\noindent
where $N$ is the number of classes (i.e. pixel labels for the segmentation task), $t_n$ is the true probability  and $y_n$ the predicted probability for class $n$. The nnU-Net employs the sum of DSC and CE as the segmentation loss function:
\begin{equation}
  \mathcal{L}_{seg} = \mathcal{L}_{\text{DSC}}  + \mathcal{L}_{\text{CE}}\,.
  \label{eq:dsc_cross_entropy}
\end{equation}

The default nnU-Net trainer uses patching into $128\times 128 \times 128$ chunks and implements a comprehensive data augmentation consisting of a sequence of input/output image/label manipulation steps. Each step is applied with random parameters according to predefined probability in sequence of: \textbf{(i)} spatial transformation by image/label rotation and scaling, \textbf{(ii)} addition of Gaussian noise, \textbf{(iii)} Gaussian blurring, \textbf{(iv)} brightness, \textbf{(v)} contrast and \textbf{(vi)} gamma intensity manipulation on the input image, \textbf{(vii)} low resolution (thick slice) simulation of the input image and \textbf{(viii)} input/output patch mirroring applied symmetrically to the input/output image/label patches.

During inference time the nnU-Net ensembles predicted segmentation probability maps of original and mirrored image patches via averaging. The comprehensive augmentation strategy and the ensembling render the nnU-Net a robust state-of-the-art deep learning model for segmentation tasks.

\subsection{nnU-Net with multi-stage unlearning}\label{subsection:proposed-method}

In multi-stage unlearning, the output of each encoder stage (\textit{E1--6}) was fed to corresponding domain classifier (blocks \textit{C1--6}).
Each classifier $Cx$ consisted of $x-1$ blocks of 3$\times$3$\times$3 convolution layer with stride 2, IN and lReLU, to achieve feature map size of $4\times 4 \times 4$. This was superseded by two Conv blocks, flattening and a fully connected layer and $N_d$ outputs corresponding to the number of \textit{seen} domains. 
Each classifier was trained by minimizing categorical CE loss as in 
Eq.~(\ref{eq:cross_entropy}), with $N$ equal to $N_d$, which is the number of domains seen during training.

The final model included one $C$ block for each $E$ stage (6 pairs in total), with the $C$ blocks trained specifically for features output by corresponding $E$ stage. For domain unlearning purposes we computed \textit{confusion loss} based on output of each $C$ block as the Kullback-Leibler divergence:
\begin{equation}
  \mathcal{L}_{conf} = -\sum_{i=1}^{N} p_i \log \left(\frac{p_i}{q_i}\right)
  \label{eq:kldivergence}
\end{equation}
\noindent
where $p_i \in P$ are the domain classifier's posterior probabilities and $q_i \in Q$ are the target uniform probabilities of domain classes. Ideally, $q_i=1/N_d$, such that noninformative predictions (eg. $p_0/p_1 = 0.5/0.5$ in case of 2 domains) yield a minimal, while highly informative predictions (e.g. $p_0/p_1 = 1.0/0.0$ in case of 2 domains) a maximal value of \textit{confusion loss}. In this way, an informative classifier output that identifies the source image domain beyond random choice allows to employ task/domain feature disentanglement and subsequent suppression of domain-specific features. 
Namely, we backpropagate $\mathcal{L}_{conf}$ through corresponding and preceding encoders, forcing them to suppress domain-specific features. As the classifier predictions become noninformative, we assume that domain information was suppressed by the encoder. This \textit{unlearning} process has to balance with the main segmentation \textit{learning} task,
for effective task/domain feature disentanglement.

The learning/unlearning schedule is detailed in Algorithm 1. 
First, during \textit{warm-up stage} (denoted by number of epochs $e_w$), the main task training takes place, in which: \textit{(1.)} the model is updated for the main segmentation task by minimizing the $\mathcal{L}_{seg}$ (Eq.~(\ref{eq:dsc_cross_entropy})) and \textit{(2.)} all the classifiers \textit{C1--6} are trained for domain classification task. Features output by each encoder block are passed to corresponding classifier, which is trained using the $\mathcal{L}_{\text{CE}}$ (Eq.~(\ref{eq:cross_entropy})).

In the subsequent \textit{learning-unlearning stage} steps \textit{(1.)} and \textit{(2.)} were followed by \textit{(3.)} unlearning of encoder blocks by backpropagation of $\mathcal{L}_{conf}$, steming from domain classifier predictions. Before computing each $\mathcal{L}_{conf}$ for \textit{C1--6} a new forward pass had to be performed, since backpropagating $\mathcal{L}_{conf}$ causes change of encoder block's forward features. For each classifier \textit{C1--6}, the corresponding $\mathcal{L}_{conf}$ was computed based on preceding encoder block features, hence, the $\mathcal{L}_{conf}$ was backpropagated via all the preceding encoder blocks. As such, the domain classifier at the U-net bottleneck, will update all \textit{E1--6} blocks of the encoder.

\begin{table}[h]
  \footnotesize
  \label{alg:unlearning-algorithm}
  \centering
  \begin{tabularx}{0.5\columnwidth}{lX}
    \multicolumn{2}{l}{\textbf{Algorithm 1:} Multi-Stage Self-Supervised Unlearning.} \\
    \midrule
    $e$ & $\gets$ \text{Current epoch} \\
    $e_t$ & $\gets$ \text{Total epochs} \\
    $e_w$ & $\gets$ \text{Warm-up epochs} \\
    $E_x$ & $\gets$ \text{Encoder block $x = \{1,\ldots,6\}$} \\
    $C_x$ & $\gets$ \text{Classifier $x = \{1,\ldots,6\}$} \\
    $Pat$ & $\gets$ \text{Patience (on learning steps)} \\
    $Tol$ & $\gets$ \text{Tolerance (on accuracy)} \\
    $UBA$ & $\gets$ \text{Upper bound accuracy} \\
    $Iter$ & $\gets$ \text{Current iteration} \\
    $Acc$ & $\gets$ \text{Classifier accuracy} \\
    ${Iter}_{UBA}$ & $\gets$ \text{Number of iterations where $Acc$ > $UBA$} \\
    $\mathcal{L}_{seg}$ & $\gets$ \text{Segmentation loss (Eq. \ref{eq:dsc_cross_entropy})} \\
    $\mathcal{L}_{CE}$ & $\gets$ \text{Classifier loss (Eq. \ref{eq:cross_entropy})} \\
    $\mathcal{L}_{conf}$ & $\gets$ \text{Confusion loss (Eq. \ref{eq:kldivergence}), for unlearning} \\
    \midrule
    \multicolumn{2}{l}{\textcolor{lightgray}{\textit{01.}} \textbf{While} $e \leq e_t$ \textbf{do}  } \\
    \multicolumn{2}{l}{\textcolor{lightgray}{\textit{02.}} \ \ \textbf{If} $e \leq e_w$ \textbf{then} \hspace{1em} } \\
    \multicolumn{2}{l}{\textcolor{lightgray}{\textit{03.}} \ \ \ \ \textcolor{lightgray}{\textit{// Main task learning }}  } \\
    \multicolumn{2}{l}{\textcolor{lightgray}{\textit{04.}} \ \ \ \ Forward pass image $\to$ Output segmentation  } \\    
    \multicolumn{2}{l}{\textcolor{lightgray}{\textit{05.}} \ \ \ \ Compute $\mathcal{L}_{seg}$ $\to$ Backpropagate $\mathcal{L}_{seg}$} \\    
    \multicolumn{2}{l}{\textcolor{lightgray}{\textit{07.}} \ \ \ \textcolor{lightgray}{\textit{// Domain classifier learning }}  } \\
    \multicolumn{2}{l}{\textcolor{lightgray}{\textit{08.}} \ \ \ \ \textbf{For all} $E_i, C_i \in E_x, C_x$; $x = \{1,\ldots,6\}$ \textbf{do}  } \\
    \multicolumn{2}{l}{\textcolor{lightgray}{\textit{09.}} \ \ \ \ \ \ Forward pass $E_i$ features $\to$ $C_i$  } \\
    \multicolumn{2}{l}{\textcolor{lightgray}{\textit{10.}} \ \ \ \ \ \ Predicted domain at stage $i \to$ compute $\mathcal{L}_{CE}^i$  } \\
    \multicolumn{2}{l}{\textcolor{lightgray}{\textit{11.}} \ \ \ \ \ \ Backpropagate $\mathcal{L}_{CE}^i \to$ to $C_i$  } \\
    \multicolumn{2}{l}{\textcolor{lightgray}{\textit{12.}} \ \ \ \ \textbf{End For}  } \\
    \multicolumn{2}{l}{\textcolor{lightgray}{\textit{13.}} \ \ \textbf{Else} \hspace{5.5em}  } \\
    \multicolumn{2}{l}{\textcolor{lightgray}{\textit{14.}} \ \ \ \ \textcolor{lightgray}{\textit{// Main task learning }}  } \\
    \multicolumn{2}{l}{\textcolor{lightgray}{\textit{15.}} \ \ \ \ Forward pass image $\to$ Output segmentation  } \\
    \multicolumn{2}{l}{\textcolor{lightgray}{\textit{16.}} \ \ \ \ Compute $\mathcal{L}_{seg}$ $\to$ Backpropagate $\mathcal{L}_{seg}$  } \\
    \multicolumn{2}{l}{\textcolor{lightgray}{\textit{18.}} \ \ \ \ \textcolor{lightgray}{\textit{// Domain classifier learning }}  } \\
    \multicolumn{2}{l}{\textcolor{lightgray}{\textit{19.}} \ \ \ \ \textbf{For all} $E_i, C_i \in E_x, C_x$; $x = \{1,\ldots,6\}$ \textbf{do}  } \\
    \multicolumn{2}{l}{\textcolor{lightgray}{\textit{20.}} \ \ \ \ \ \ Forward pass $E_i$ features $\to$ $C_i$  } \\
    \multicolumn{2}{l}{\textcolor{lightgray}{\textit{21.}} \ \ \ \ \ \ Predicted domain for stage $i \to$ compute $\mathcal{L}_{CE}^i$, ${Acc}^i$  } \\
    \multicolumn{2}{l}{\textcolor{lightgray}{\textit{22.}} \ \ \ \ \ \ Backpropagate $\mathcal{L}_{CE}^i \to$ to $C_i$  } \\
    \multicolumn{2}{l}{\textcolor{lightgray}{\textit{23.}} \ \ \ \ \textbf{End For}  } \\
    \multicolumn{2}{l}{\textcolor{lightgray}{\textit{24.}} \ \ \ \ \textbf{For all} $E_i, C_i \in E_x, C_x$; $x = \{1,\ldots,6\}$ \textbf{do}  } \\
    \multicolumn{2}{l}{\textcolor{lightgray}{\textit{25.}} \ \ \ \ \ \ \textcolor{lightgray}{\textit{// Self-supervision }}  } \\
    \multicolumn{2}{l}{\textcolor{lightgray}{\textit{26.}} \ \ \ \ \ \ \textbf{If} $Acc^i > UBA$   } \\
    \multicolumn{2}{l}{\textcolor{lightgray}{\textit{27.}} \ \ \ \ \ \ \ \ ${Iter}^{\,i}_{UBA} += 1$          }\\
    \multicolumn{2}{l}{\textcolor{lightgray}{\textit{28.}} \ \ \ \ \ \ \textbf{Else}        }\\
    \multicolumn{2}{l}{\textcolor{lightgray}{\textit{29.}} \ \ \ \ \ \ \ \ ${Iter}^{\,i}_{UBA} = 0$          }\\
    \multicolumn{2}{l}{\textcolor{lightgray}{\textit{30.}} \ \ \ \ \ \ \textbf{End If}        }\\
    \multicolumn{2}{l}{\textcolor{lightgray}{\textit{31.}} \ \ \ \ \ \ \textcolor{lightgray}{\textit{// Unlearning }}  } \\
    \multicolumn{2}{l}{\textcolor{lightgray}{\textit{32.}} \ \ \ \ \ \ \textbf{If} ${Iter}^{\,i}_{UBA} > Pat$     }\\
    \multicolumn{2}{l}{\textcolor{lightgray}{\textit{34.}} \ \ \ \ \ \ \ \ Forward pass $E_i$ features $\to$ to $C_i$  } \\
    \multicolumn{2}{l}{\textcolor{lightgray}{\textit{35.}} \ \ \ \ \ \ \ \ Predicted domain for stage $i \to$ compute $\mathcal{L}_{conf}^i$  } \\
    \multicolumn{2}{l}{\textcolor{lightgray}{\textit{36.}} \ \ \ \ \ \ \ \ Backpropagate $\mathcal{L}_{conf}^i \to$ $E_{x,\ldots, 1}$ }\\
    \multicolumn{2}{l}{\textcolor{lightgray}{\textit{37.}} \ \ \ \ \textbf{End For}  } \\
    \multicolumn{2}{l}{\textcolor{lightgray}{\textit{38.}} \textbf{End While}  } \\
    \bottomrule
  \end{tabularx}
\end{table}

In each iteration we form a batch of equal number of images for each source domain. For instance, a batch of 8 input images from 2 source domains included 4 images per each domain.
Providing unequal number of image per domain impedes model unlearning ability, since the confusion loss function (Eq. \ref{eq:kldivergence}) optimizes model towards non-informative predictions, which are best discernible at equal source domain distributions.

\subsection{Self-supervised learning–unlearning schedule}
\label{sec:selfsupervised_schedule}

    Following the self-configuring philosophy of nnU-Net, we employ a self-supervised \emph{learning–unlearning} schedule.
It is grounded in prior domain-adversarial and domain-confusion work ~\cite{ganin_domain-adversarial_2016,tzeng_deep_2014}, and information-bottleneck theory suggesting that task-agnostic MRI scanner-specific nuisance signals reside in higher-level representations~\cite{tishby_deep_2015}, rendering unlearning effective. However, in prior study MRI unlearning from bottleneck layer ~\cite{dinsdale2021unlearning} was rather ineffective, despite hyperparameter tuning. This necessitates curriculum-style patience mechanism~\cite{soviany_curriculum_2022} that triggers training changes only when a stable above-chance domain signal is detected.

Concretely, for a domain classifier $C_i$ operating at a given encoder depth we compute an \emph{upper-bound accuracy} (UBA) as:
\begin{equation}
  \text{UBA} \;=\; \frac{1}{N_d} + Tol,\qquad Tol\in[0,1-\frac{1}{N_d}],
  \label{eq:uba}
\end{equation}
where $N_d$ is the number of domains (e.g. scanners) and $Tol$ is a small tolerance above random chance (0.5) that indicates a practically meaningful level of domain discriminability. While monitoring accuracy $Acc^i_t$ of classifier $C_i$ at iteration $t$  a counter $Iter_{UBA}^i$ of consecutive iterations where $Acc^i_t \!>\! \text{UBA}$ is maintained. When $Iter_{UBA}^i$ reaches the patience threshold $Pat$ the training alternates to an unlearning (gradient reversal) step for the particular classifier. Specifically, the confusion loss $\mathcal{L}^i_{\text{conf}}$ is backpropagated via corresponding and previous encoders ($E_j;\,j=1,\ldots,i$) in order to force the classifier toward random chance accuracy. When in effect, unlearning is applied alternately with the segmentation loss $\mathcal{L}^i_{seg}$ in 
to preserve segmentation performance.

\section{Experiments \& Results}

The proposed and five benchmark unsupervised domain adaptation models and/or strategies were applied on the WMH's seen train subset (Table~\ref{table:wmh-data-description}), using 32 cases for training and 8 for validation and hyperparameter tuning. The five benchmarks were: \textbf{(i)} the nnU-Net without data augmentation; \textbf{(ii)} the default nnU-Net; \textbf{(iii)} the default nnU-Net applied to N4 bias corrected and denoised train and test subsets; and \textbf{(iv)} a \cite{dinsdale2021unlearning} unlearning approach implemented into nnU-Net. Weight initialization and order of training data were randomized, using fixed seed. %

Ablation study involved evaluation of the proposed approach on WMH's unseen subset of 50 scans from three scanners. We systematically explored unlearning strategies and schedules, evaluated their impact on the WML segmentation performance and generalization capability, and determined the best model. To avoid bias, these scans were not further used for evaluation purposes.

Comparative performance analysis involved all five benchmark and the best proposed approach as determined in the ablation study. Quantitative evaluation was performed on three unseen test datasets (MSSEG, MSLJ and ISBI) on a total of 103 FLAIR scans. Additionally, evaluation was performed on 60 test FLAIR scans in the \textit{seen} WMH dataset (different subjects, but same scanners as in training) to comparatively assess the impact of scanner variability.

\subsection{Evaluation metrics}\label{subsection:evaluation-metrics}

Output segmentation maps were evaluated using established metrics, such as the DSC ($1-\mathcal{L}_{\text{DSC}}$, cf. Eq.~(\ref{eq:dsc})) and the TP Rate (TPR) defined as:
\begin{equation}
  \text{TPR} = \frac{\text{TP}}{\text{TP} + \text{FN}}\,.
  \label{eq:tpr}
\end{equation}

We further employed lesion-specific metrics, such as lesion-wise TPR (LTPR) (\ref{eq:ltpr}) and lesion-wise False Discovery Rate (LFDR) (\ref{eq:lfdr}) with predefined \textit{intersection-over-union threshold} at $t_{IOU} = 0.05$. Specifically, the LTPR was defined as the ratio between the number of predicted lesions that overlap with lesions in the ground-truth, with percentage overlap higher than $t_{IOU}$, and the total number of lesions in the ground-truth, namely:
\begin{equation}
  \text{LTPR} = \frac{\text{\#(LTP)}}{\text{\#(RL)}}\,,
  \label{eq:ltpr}
\end{equation}
\noindent
where LTP stands for \textit{Lesion True Positives} and RL for \textit{Reference Lesions}, and operator $\#(\cdot)$ denotes \textit{lesion count}, obtained via 18-connected neighborhood labeling operator.

The LFDR was defined as the ratio between number of predicted lesions that do not overlap with any lesions in the ground-truth (overlap less than $t_{IOU}$), and the total number of predicted lesions and computed as:
\begin{equation}
  \text{LFDR} = \frac{\text{\#(LFP)}}{\text{\#(PL)}}\,,
  \label{eq:lfdr}
\end{equation}
\noindent
where LFP stands for \textit{Lesion False Positives} and PL for \textit{Predicted Lesions}.

Volumetric discrepancy between predicted and ground-truth segmentations was evaluated by the Relative Volume Error (RVE), defined as follows:
\begin{equation}
  \text{RVE} = \frac{|V(PL) - V(RL)|}{V(RL)}\,,
  \label{eq:RVE}
\end{equation}
\noindent
where 
$V(L)$ represents the volume obtained via pixel count of the specified label $L$.

To compare the overall performance in a set of different methods (or models or strategies) accounting for all aforementioned metrics we used an average rank score defined as:
\begin{equation}
\begin{split}
RS  = & \frac{1}{5}\cdot\Bigr[ R(\text{DSC}) + R(\text{TPR}) + R(\text{LTPR})\\
&  + R(\text{LFDR}) + R(\text{RVE})\Bigr]\,,
\end{split}
  \label{eq:rs}
\end{equation}
\noindent
where $R(m)$ is the ranking of the method in the set according to  particular metric $m$. When having $M$ methods in comparison, RS is in range [1.0, $M$].

To analyze the effectiveness of unlearning we also compute for each domain classifier $C_i$ its accuracy as $Acc^i=TP+TN/(TP+TN+FP+FN)$.

In results we report mean ($\mu$) and standard deviation (SD), or, where indicated, the bootstrap 95\% confidence interval (CI) for all evaluated metrics. In performance comparisons we used Wilcoxon signed-rank test \citep{wilcoxon_individual_1945}, or, where applicable, independent sample t-test, to compare metric values of our proposed strategy against each of the other tested models or strategies. To adjust for multiple pairwise comparisons we used Bonferroni correction by multiplying  the obtained \textit{p-values} by the number of methods in comparison. If the \textit{adjusted p-value} was less than the significance threshold $\alpha=0.05$, we considered the observed difference statistically significant.

\subsection{Ablation study}
\label{subsection:ablation-study}

Results in Table~\ref{table:ablation-study} provide a comparison of effectiveness of unlearning strategies, exploring the hyperparameter space by varying: \textbf{(i)} stages that are unlearned (indicated by checkmarks and corresponding to stages as depicted in Figure \ref{figure:architecture}) 
and \textbf{(ii)} the ratio of learning to unlearning training steps (LUR); for instance, setting LUR to 5-1 indicates a schedule of 5 learning iterations of main task and 1 iteration of encoder unlearning.
In case of using \textit{self-supervised} learning-unlearning strategy as described in Section \ref{subsection:proposed-method}, we simply indicated specific stages that \textit{were} being unlearned, but for which the proposed strategy controlled the execution of unlearning steps as described in section \ref{sec:selfsupervised_schedule}. 

\begin{figure}[!b]
  \centering
  \includegraphics[width=0.75\textwidth]{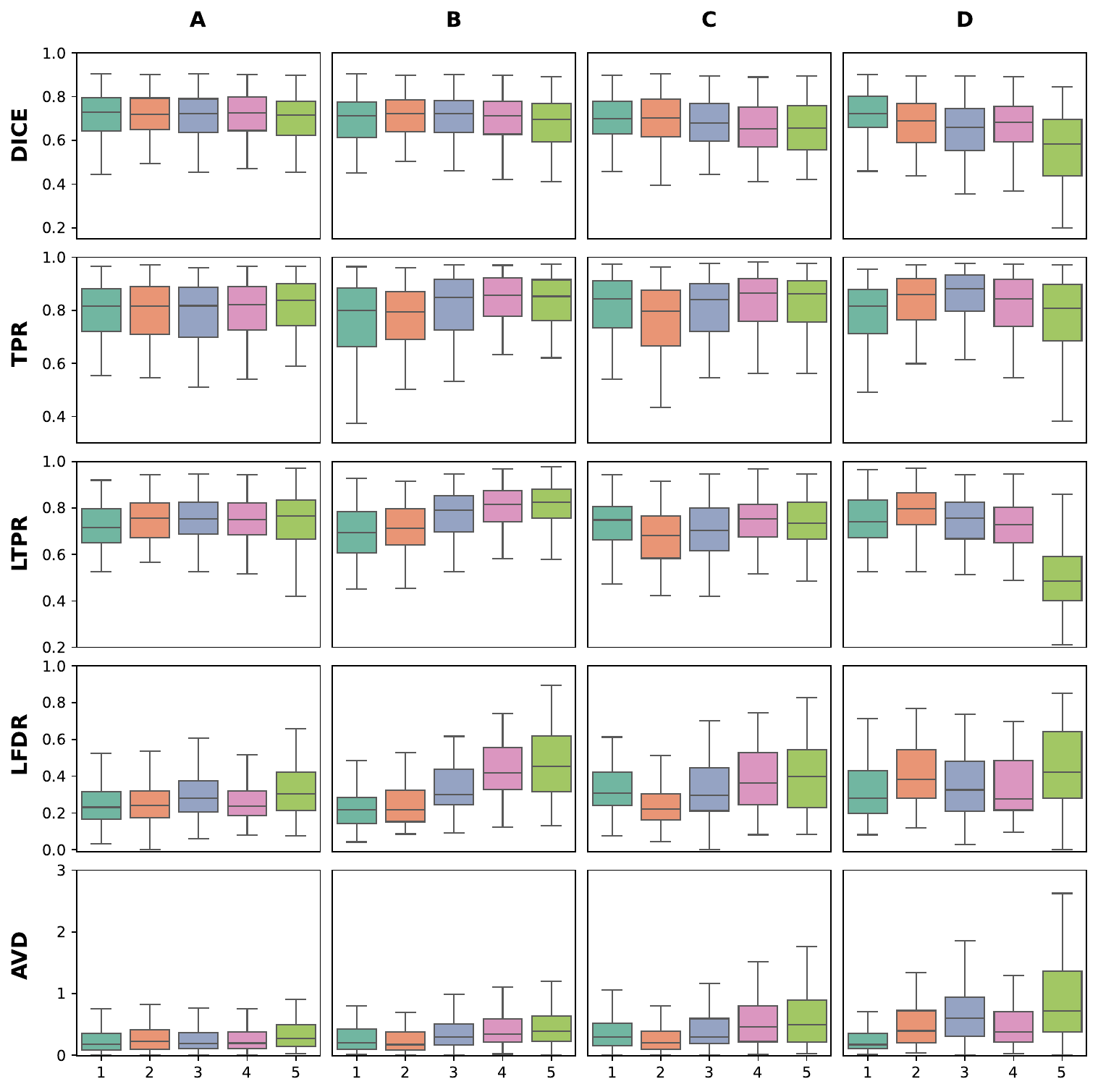}
  \caption{Ablation study metric values with respect to learn-to-unlearn ratio (LUR=1-1, 2-1, 3-1, 4-1, 5-1), for various unlearning stages: \textit{A - from bottleneck, B - bottom three blocks, C - top three blocks, D - entire encoder}.} \label{figure:ablation-metrics-stages-boxplots}
\end{figure}

We have tested several combinations of LUR setting and of unlearning stages, but listed in Table~\ref{table:ablation-study} only those that achieved best value per each evaluated performance metric. 
All results are visualized in Figure~\ref{figure:ablation-metrics-stages-boxplots}. 
Establishing the optimal set of unlearning stages and setting LUR is rather challenging, since for various settings different segmentation performance metrics attain best values (bolded values in Table~\ref{table:ablation-study}). While the self-supervised model does not achieve the best value in any performance metric,  
it provided the most balanced overall performance across all metrics. Namely, the average ranking metric RS highlighted the best overall performance of the self-supervised model, specifically the one with bottom 3 unlearning stages, scoring best RS of 4.0. In subsequent experiments we used this model, dubbed self-supervised multi-stage unlearning (SSMSU) model.

\begin{table*}[bthp]
  \centering
  \footnotesize
  \caption{Ablation study results on the 2017 WMH Segmentation Challenge's unseen test split. Values are mean $\pm$ standard deviation, while \textbf{bold} indicates best result for particular evaluation metric. \textbf{LUR} is the ratio of the number of learning vs. unlearning steps.}
  \label{table:ablation-study}
  \begin{tabular}{|cccccc|c|c|c|c|c|c|c|}
      \hline
      \multicolumn{5}{|r}{\textbf{Unlearning stages}} & & 
      \multirow[b]{2}{*}{\textbf{LUR}} & \multicolumn{6}{c|}{\textbf{Unseen data} - \textit{not used for model training: Philips 3T 3D, GE 3T 3D, GE 1.5T 3D}} \\
      \cline{1-6}
      \cline{8-13}
      1 & 2 & 3 & 4 & 5 & 6 &  & \textbf{DSC}$\uparrow$ & \textbf{TPR}$\uparrow$ & \textbf{LTPR}$\uparrow$ & \textbf{LFDR}$\downarrow$ & \textbf{RVE}$\downarrow$ & \textbf{RS}$\downarrow$ \\
      \toprule

      \bottomrule
       &            &            &            &            & \checkmark & 1 - 1                   & \textbf{0.719 ± 0.100} & 0.794 ± 0.116 & 0.717 ± 0.097 & 0.251 ± 0.113 & 0.262 ± 0.235          & 4.2 \\
       &  &  & \checkmark & \checkmark &  \checkmark & 2 - 1                   & 0.705 ± 0.103 & 0.767 ± 0.137 & 0.716 ± 0.107 & \textbf{0.243 ± 0.115} & \textbf{0.241 ± 0.201} & 5.0 \\
      \checkmark & \checkmark & \checkmark & \checkmark & \checkmark & \checkmark & 3 - 1                   & 0.642 ± 0.132 & \textbf{0.856 ± 0.096} & 0.741 ± 0.104 & 0.347 ± 0.185 & 0.763 ± 0.589          & 6.4 \\
       & & & \checkmark & \checkmark & \checkmark & 5 - 1                   & 0.674 ± 0.120 & 0.832 ± 0.101 & \textbf{0.813} ± 0.095 & 0.474 ± 0.185 & 0.527 ± 0.449          & 5.2 \\
      \toprule

      \bottomrule
       &            &            &            &            & \checkmark & \multirow{4}{*}{\rotatebox[origin=r]{90}{Self-superv.}} & 0.712 ± 0.099 & 0.805 ± 0.115 & 0.759 ± 0.103 & 0.294 ± 0.136 & 0.308 ± 0.272 & 4.4 \\
       & &            &            &        \checkmark    & \checkmark &                                                     & 0.706 ± 0.103 & 0.809 ± 0.126 & 0.764 ± 0.102 & 0.338 ± 0.134 & 0.335 ± 0.254 & 5.0 \\
       &  &  &       \checkmark     &       \checkmark     & \checkmark &                                                     & 0.715 ± 0.105 & 0.814 ± 0.120 & 0.764 ± 0.095 & 0.313 ± 0.136 & 0.316 ± 0.250 & \textbf{4.0} \\
       &  & \checkmark & \checkmark &      \checkmark      & \checkmark &                                                     & 0.697 ± 0.112 & 0.829 ± 0.117 & 0.777 ± 0.100 & 0.310 ± 0.134 & 0.416 ± 0.338 & 4.4 \\
       & \checkmark & \checkmark & \checkmark & \checkmark & \checkmark &                                                     & 0.633 ± 0.125 & 0.817 ± 0.120 & 0.785 ± 0.109 & 0.446 ± 0.169 & 0.664 ± 0.557 & 6.2 \\
      \toprule
    \end{tabular}
\end{table*}

Results in Figure~\ref{figure:ablation-metrics-stages-boxplots} show that
\textbf{(i)} most stage settings exhibited better performance with balanced LUR of 1-1, while LUR of 5-1 proved to be worst across most metrics and stages. An exception are TPR and LTPR for setting \textit{B}, but which has very high LFDR of 0.527.
\textbf{(ii)} We notice overall poor performance in setting \textit{D}, with lowest values for all metrics at LUR $5-1$. As expected \textbf{(iii)} with increasing the LUR ratio the TPR and LTPR tend to increase (cf. settings \textit{A}, \textit{B} and \textit{C}), since performing more learning vs. unlearning steps benefits the main segmentation task. However, \textbf{(iv)} there are higher variations in performance for settings \textit{C} and \textit{D}, whereas \textit{A} and \textit{B} have consistent result across all LURs. 

Table \ref{table:domain-accuracies-top-bot} shows
the nnU-Net and the SSMSU post warm-up stage was clearly able to identify with 100\% accuracy the domain at all six stages of the encoder. After unlearning the upper 3 stages, the accuracy of domain classifier in these 3 stages were in range 77-85\%, which is rather high and indicates poor unlearning capability. When unlearning bottom 3 stages, the corresponding classifier accuracies were 48-67\%, and 48-50\% for bottom 2 stages ($i=5,6$). Interestingly, when unlearning all stages the unlearning was generally less effective, leading to convergence only at the bottleneck ($i=$1--5: 99--65\%; $i=6$: 52\%). Unlearning was effective only in the deeper encoder stages because these capture higher-level feature representations, in which scanner-specific artifacts seem more entangled with the lesion appearance, making them the primary locus where removing scanner information yields measurable gains with respect to the segmentation task.

\begin{table*}[!b]
  \caption{Domain classifier accuracy, post warm-up and post-unlearning, per stage of the nnU-Net SSMSU on 2017 WMH dataset's seen train split, comparing top- and bottom-three and unlearning of all stages. Values in \textbf{bold} indicate stages actively unlearned.}
  \label{table:domain-accuracies-top-bot}
  \raggedright
  \begin{tabular}{|l|l|c|c|c|c|c|c|}
    \hline
    \textbf{Model} &  ~ &  \multicolumn{6}{c|}{\textbf{Classifier accuracy, $Acc^i$}} \\ \cline{3-8}
    Unlearning stages   & Step &  \textbf{i=1} & \textbf{i=2} & \textbf{i=3} & \textbf{i=4} & \textbf{i=5} & \textbf{i=6} \\ \hline\hline
    nnU-Net*            & ~       & 100\% & 100\% & 100\% & 100\% & 100\% & 100\% \\ 
    \hline\hline
    nnU-Net* SSMSU      & Post warm-up & 100\% & 100\% & 100\% & 100\% & 100\% & 99\% \\ \cline{2-8}
    Unlearning stages: 1 \& 2 \& 3 (Top 3) & Post unlearning & \textbf{85\%} & \textbf{77\%} & \textbf{80\%} & 98\% & 99\% & 91\% \\ 
    \hline\hline
    nnU-Net* SSMSU      & Post warm-up & 100\% & 99\% & 99\% & 99\% & 99\% & 99\% \\ \cline{2-8}
    Unlearning stages: 4 \& 5 \& 6 (Bottom 3) & Post unlearning     & 100\% & 100\% & 100\% & \textbf{67\%} & \textbf{48\%} & \textbf{50\%} \\
    \hline\hline
    nnU-Net* SSMSU      & Post warm-up & 100\% & 99\% & 99\% & 99\% & 99\% & 99\% \\ \cline{2-8}
    Unlearning stages: 1 -- 6 (All) & Post unlearning     & \textbf{99\%} & \textbf{90\%} & \textbf{76\%} & \textbf{65\%} & \textbf{67\%} & \textbf{52\%} \\ \hline
    \multicolumn{8}{l}{\footnotesize *Predefined default trainer in nnU-Net.}
  \end{tabular}
\end{table*}

\subsection{Lesion segmentation}

Lesion segmentation results for the proposed and other models and domain adaptation strategies are shown in Table \ref{table:lesion-segmentation-results-comparison}. 
The WMH results on the seen subset indicate that values of DSC and RVE for the proposed SSMSU model are on par with the default nnU-Net. On the other hand, the proposed model comparatively improved TPR and LTPR, while LFDR increased (all noted differences were statistically significant). Despite a larger LFDR (Lesion False Detection Rate), indicating more incorrectly detected lesions. However, these were small FP lesions as indicated by the lower value of RVE for the proposed model. 

\begin{figure*}
  \centering
  \includegraphics[width=1.0\textwidth]{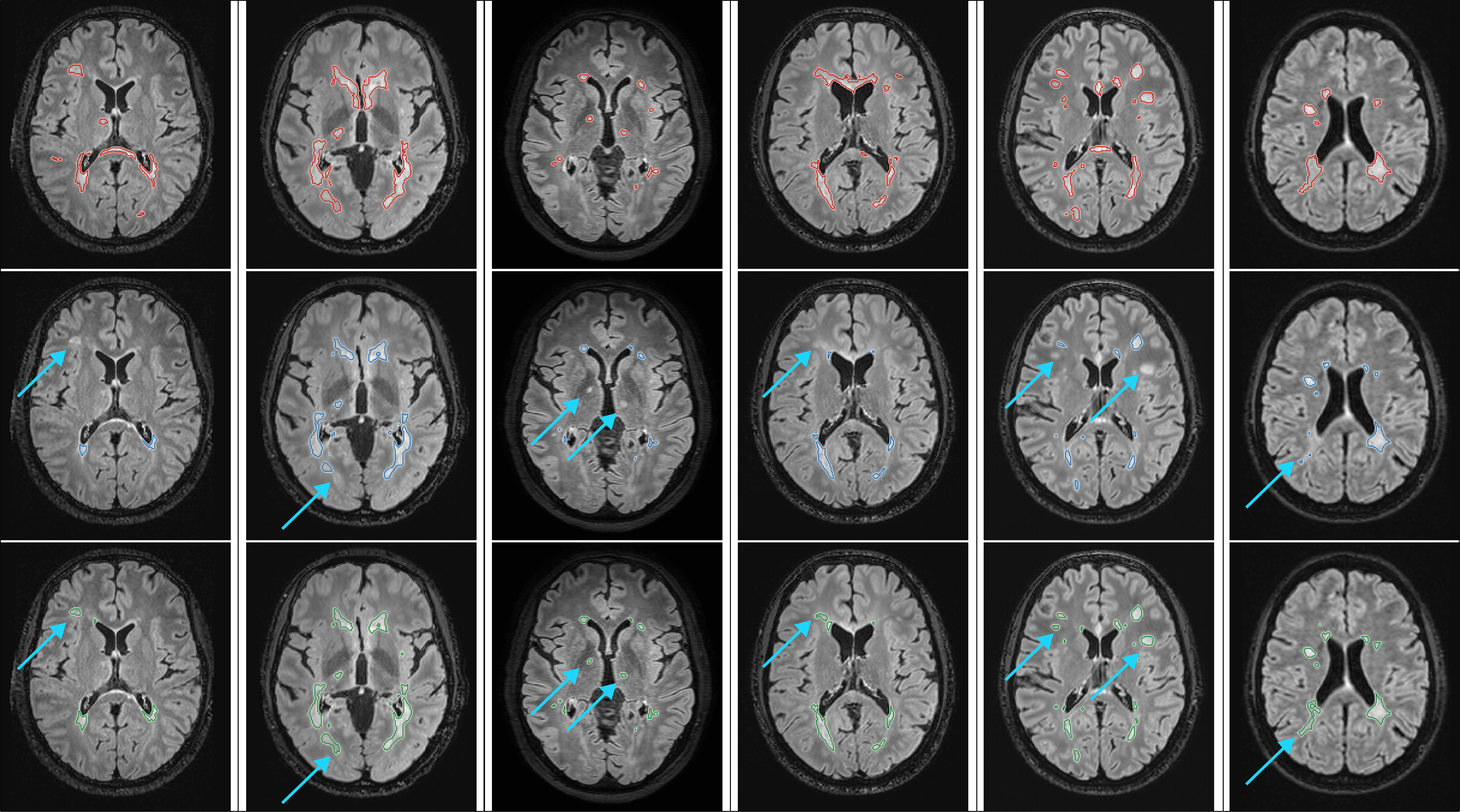}
  \caption{Axial cross-sections of MSSEG 2016 MR scans with superimposed manual (\textit{top row}) and second best default nnU-Net (\textit{middle row}) and the proposed SSMSU model segmentations (\textit{bottom row}). \textit{Arrows} highlight important differences.} \label{figure:qc-m16-join}
\end{figure*}

When applying models to the unseen scanner data in MSSEG, MSLJ and ISBI datasets, we observe a larger gap between methods, with DSC scores ranging from 0.52--0.64, 0.48--0.57 and 0.54--0.67, respectively. The proposed SSMSU outperformed other tested models, with significant increases in DSC, TPR, LTPR (except in one comparison to Dinsdale et al.), and significantly lower RVE metric value. Similarly as on the seen dataset, the LFDR was significantly higher compared to all other model results, but these FPs were small based on small RVE values. 
Figure \ref{figure:qc-m16-join} visualizes axial MRI cross-sections with superimposed manual, nnU-Net and SSMSU segmentations; the highlights clearly indicate the higher sensitivity to lesions of the SSMSU model.

\begin{figure}
  \includegraphics[width=0.75\textwidth]{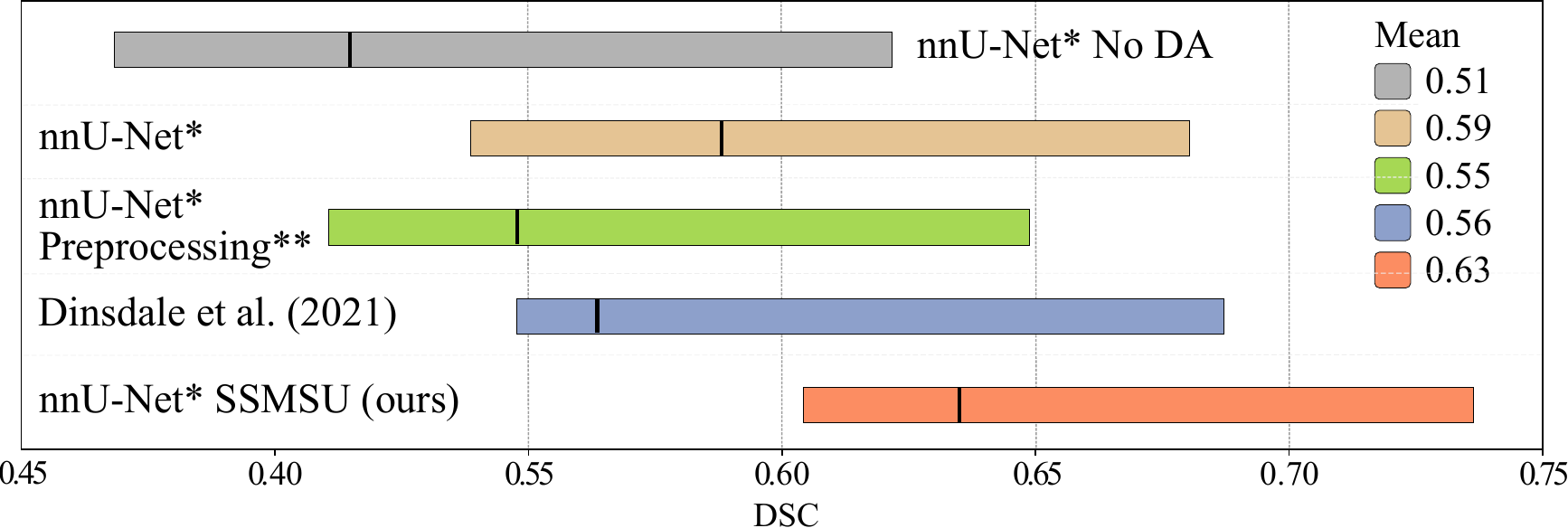}
  \vspace{-0.3cm}
  \caption{DSC with 95\% confidence intervals for MSSEG 2016 unseen test set. For notes (*,**) see caption of Table~\ref{table:lesion-segmentation-results-comparison}.}
  \label{figure:ci-m16-dice}
\end{figure}

\begin{figure}
  \includegraphics[width=0.75\textwidth]{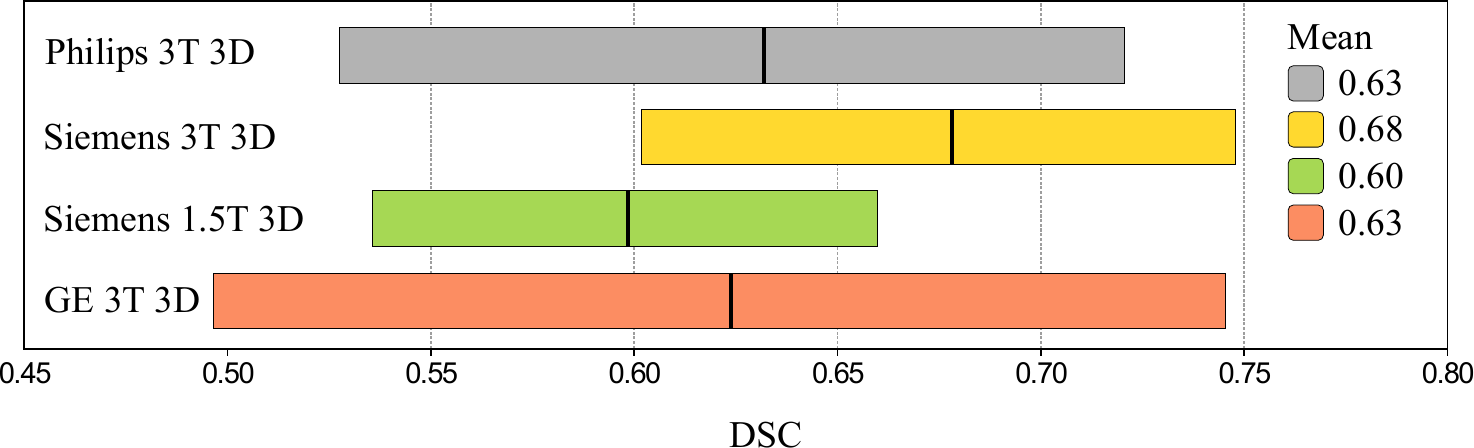}
  \caption{DSC with 95\% confidence intervals per source domain of MSSEG unseen test set for nnU-Net SSMSU. }
  \label{figure:ci-m16-dice-per-vendors}
\end{figure}

The 95\% CIs for the DSC values on MSSEG unseen test set on Figure \ref{figure:ci-m16-dice} indicate clearly improved results by the SSMSU compared to other tested methods. Furthermore, Figure \ref{figure:ci-m16-dice-per-vendors} shows the proposed strategy had consistent DSC values across different scanners in MSSEG, albeit slightly higher for the Siemens 3T 3D.

In order to better contextualize our results we conducted a comparative analysis with the LST-AI model~\cite{wiltgen_lst-ai_2024}, which was trained on 491 cases and bimodal T1-weighted and FLAIR input. Using the authors’ reported results on the same three unseen test sets, and independent t-tests reported in Table~\ref{table:lesion-segmentation-results-comparison}, we observed that: (i) segmentation performance on MSSEG was comparable between LST-AI and our FLAIR-only SSMSU model; (ii) LST-AI achieved significantly higher DSC on MSLJ; and (iii) SSMSU performed slightly better than LST-AI on ISBI, according to higher DSC. TPR and LTPR were also significantly higher for LST-AI on MSSEG and MSLJ, whereas they were comparable on ISBI. Notably, LST-AI consistently achieved lower RVE.

\begin{table*}[hbp]
  \raggedright
  \caption{Segmentation evaluation on seen (WMH) and three unseen (MSSEG, MSLJ, ISBI) datasets, using respective seen and unseen test splits (Table~\ref{table:wmh-data-description}). Values are mean $\pm$ standard deviation, while \textbf{bold} indicates best results per each performance metric for methods in direct comparison.}
  \label{table:lesion-segmentation-results-comparison}
  \begin{tabular}{|l|c|c|c|c|c|c|}
  \multicolumn{7}{l}{\textbf{WMH - \textit{seen test dataset}}} \\
  \hline
  \textbf{Model}                & \textbf{DSC}$\uparrow$                   & \textbf{TPR}$\uparrow$            & \textbf{LTPR}$\uparrow$          & \textbf{LFDR}$\downarrow$            & \textbf{RVE}$\downarrow$          & \textbf{RS}$\downarrow$ \\   \hline\hline
  nnU-Net No DA                         & 0.750 ± 0.120$^\dag$                 & 0.775 ± 0.132$^\dag$          & 0.680 ± 0.160$^\dag$          & 0.390 ± 0.160$^\dag$            & 0.220 ± 0.353       & 3.2  \\
  nnU-Net*                      & \textbf{0.773 ± 0.108} & 0.785 ± 0.129$^\dag$          & 0.679 ± 0.103$^\dag$          & \textbf{0.210 ± 0.131}   & 0.205 ± 0.388       & 2.0   \\
  nnU-Net* Preprocessing**    & 0.732 ± 0.123$^\dag$                 & 0.748 ± 0.149$^\dag$          & 0.630 ± 0.105$^\dag$          & 0.219 ± 0.137$^\dag$            & 0.232 ± 0.371       & 3.8  \\
  Dinsdale et al.~\cite{dinsdale2021unlearning}               & 0.724 ± 0.149$^\dag$                 & 0.793 ± 0.123$^\dag$          & 0.624 ± 0.108$^\dag$          & 0.426 ± 0.187$^\dag$            & 0.304 ± 0.477       & 4.4  \\
  nnU-Net SSMSU (ours)   & 0.772 ± 0.097                 & \textbf{0.810 ± 0.124} & \textbf{0.718 ± 0.097} & 0.262 ± 0.160             & \textbf{0.197 ± 0.228}     & \textbf{1.6}    \\
  \hline
\multicolumn{6}{l}{}\\[-6pt]
\multicolumn{6}{l}{\textbf{MSSEG - \textit{unseen test dataset}}} \\
  \hline
  \textbf{Model}                  & \textbf{DSC}$\uparrow$            & \textbf{TPR}$\uparrow$            & \textbf{LTPR}$\uparrow$           & \textbf{LFDR}$\downarrow$          & \textbf{RVE}$\downarrow$         & \textbf{RS}$\downarrow$ \\  \hline\hline
  nnU-Net No DA                           & 0.515 ± 0.198$^\dag$           & 0.468 ± 0.145$^\dag$           & 0.628 ± 0.188$^\dag$           & 0.684 ± 0.227$^\dag$          & 0.877 ± 2.042$^\dag$        & 4.2 \\
  nnU-Net*                        & 0.588 ± 0.169$^\dag$           & 0.479 ± 0.163$^\dag$           & 0.619 ± 0.188$^\dag$           & \textbf{0.281 ± 0.197} & 0.516 ± 0.668$^\dag$        & 2.6 \\
  nnU-Net* Preprocessing**      & 0.548 ± 0.185$^\dag$           & 0.442 ± 0.168$^\dag$           & 0.584 ± 0.186$^\dag$           & 0.337 ± 0.229$^\dag$          & 0.493 ± 0.330$^\dag$        & 3.6 \\
  Dinsdale et al.~\cite{dinsdale2021unlearning}                 & 0.564 ± 0.204$^\dag$           & 0.548 ± 0.150    & \textbf{0.740 ± 0.179}  & 0.708 ± 0.204$^\dag$          & 0.828 ± 2.189$^\dag$        & 3.0 \\
  nnU-Net SSMSU (ours)      & \textbf{0.635 ± 0.164}  & \textbf{0.563 ± 0.156}  & 0.700 ± 0.171  & 0.384 ± 0.199          & \textbf{0.420 ± 0.785}   & \textbf{1.6}     \\
  \hline
  LST-AI*** \cite{wiltgen_lst-ai_2024}     & 0.659 ± 0.159  & 0.675 ± 0.172$^\ddag$   & 0.815 ± 0.170$^\ddag$   & /          & 0.094 ± 0.568$^\ddag$    & /     \\
  \hline   
\multicolumn{6}{l}{}\\[-6pt]
\multicolumn{6}{l}{\textbf{MSLJ - \textit{unseen test dataset}}} \\
 \hline
  \textbf{Model}                & \textbf{DSC}$\uparrow$                   & \textbf{TPR}$\uparrow$            & \textbf{LTPR}$\uparrow$          & \textbf{LFDR}$\downarrow$            & \textbf{RVE}$\downarrow$          & \textbf{RS}$\downarrow$ \\   \hline\hline
 nnU-Net No DA         & 0.475 ± 0.183$^\dag$ & 0.357 ± 0.146$^\dag$ & 0.375 ± 0.108$^\dag$ & 0.476 ± 0.262$^\dag$ & 0.519 ± 0.141$^\dag$ & 4.8  \\
 nnU-Net*      & 0.523 ± 0.194$^\dag$ & 0.403 ± 0.178$^\dag$ & 0.389 ± 0.120$^\dag$ & \textbf{0.151 ± 0.151} & 0.507 ± 0.180$^\dag$ & 2.4 \\
 nnU-Net* Preprocessing** & 0.507 ± 0.181$^\dag$ & 0.384 ± 0.156$^\dag$ & 0.431 ± 0.100$^\dag$ & 0.194 ± 0.183$^\dag$ & 0.512 ± 0.149$^\dag$ & 3.0 \\
 Dinsdale et al.~\cite{dinsdale2021unlearning}            & 0.501 ± 0.171$^\dag$ & 0.389 ± 0.137$^\dag$ & 0.401 ± 0.111$^\dag$ & 0.492 ± 0.264$^\dag$ & 0.468 ± 0.145$^\dag$ & 3.4 \\
 nnU-Net SSMSU (Ours)        & \textbf{0.574 ± 0.170} & \textbf{0.475 ± 0.161} & \textbf{0.479 ± 0.108} & 0.273 ± 0.196 & \textbf{0.371 ± 0.165} & \textbf{1.4} \\
\hline
  LST-AI*** \cite{wiltgen_lst-ai_2024}     & 0.742 ± 0.104$^\ddag$   & 0.703 ± 0.138$^\ddag$   & 0.617 ± 0.125$^\ddag$   & /          & 0.121 ± 0.153$^\ddag$    & /     \\
  \hline  
\multicolumn{6}{l}{}\\[-6pt]
\multicolumn{6}{l}{\textbf{ISBI - \textit{unseen test dataset}}} \\
\hline
  \textbf{Model}                & \textbf{DSC}$\uparrow$                   & \textbf{TPR}$\uparrow$            & \textbf{LTPR}$\uparrow$          & \textbf{LFDR}$\downarrow$            & \textbf{RVE}$\downarrow$          & \textbf{RS}$\downarrow$ \\   \hline\hline
nnU-Net No DA                         & 0.535 ± 0.149$^\dag$ & 0.398 ± 0.154$^\dag$ & 0.500 ± 0.157$^\dag$ & 0.353 ± 0.178$^\dag$ & 0.554 ± 0.169$^\dag$ & 4.6 \\
nnUnet*                      & 0.609 ± 0.137$^\dag$ & 0.477 ± 0.165$^\dag$ & 0.543 ± 0.131$^\dag$ 
& \textbf{0.154 ± 0.098} & 0.472 ± 0.190$^\dag$ & 2.2 \\
nnU-Net Preprocessing** & 0.540 ± 0.139$^\dag$ & 0.393 ± 0.143$^\dag$ & 0.574 ± 0.119 & 0.191 ± 0.115 & 0.578 ± 0.149$^\dag$ & 3.4 \\
Dinsdale et al.~\cite{dinsdale2021unlearning}                             & 0.588 ± 0.130$^\dag$ & 0.471 ± 0.150$^\dag$ & 0.522 ± 0.158$^\dag$ & 0.437 ± 0.146$^\dag$ & 0.428 ± 0.210$^\dag$ & 3.4 \\
nnU-Net SSMSU (ours)                        & \textbf{0.674 ± 0.115} & \textbf{0.560 ± 0.154} & \textbf{0.575 ± 0.138} & 0.203 ± 0.104 & \textbf{0.366 ± 0.191} & \textbf{1.4} \\
\hline
  LST-AI*** \cite{wiltgen_lst-ai_2024}     & 0.609 ± 0.126  & 0.540 ± 0.151  & 0.552 ± 0.150  & /          & 0.247 ± 0.174$^\ddag$   & /     \\
  \hline 
  \end{tabular}

  \vspace{0.1cm}
 \scriptsize
  \begin{tabular}{l}
 
    DA-Data Augmentation; SSMSU-Self-Supervised Multi-Stage Unlearning; *Predefined default trainer in nnU-Net. 
    **Preprocessing done on train and test splits with denoise\\and N4 bias correction.
    ***Inputs both T1W and FLAIR scans. Results were sourced from Supplementary Data 1 in~\citep{wiltgen_lst-ai_2024}.
    $^\dag$Mean significantly different (adj. p<0.05; Bonferroni\\correction) from \textit{nnU-Net SSMSU (ours)}, as per  Wilcoxon signed-rank test. $^\ddag$ Mean significantly different (p<0.05) from \textit{nnU-Net SSMSU (ours)}, as per independent sample t-test.
   
  \end{tabular}

\end{table*}

\section{Discussion}\label{section:discussion}

This study on unsupervised domain adaptation introduced and comparatively evaluated a novel multi-stage self-supervised unlearning strategy dubbed SSMSU in context of WML segmentation. The strategy was implemented into the state-of-the-art nnU-Net framework and the model trained on \textit{seen} train subset of WMH dataset, and evaluated on WMH's 
\textit{seen} and three other \textit{unseen} test subsets. Results on the unseen scans consistenly show that the SSMSU substantially improved WML segmentation performance compared to the other tested methods and strategies.

\subsection{Self-supervised multi-stage unlearning}

Prior work has explored domain unlearning strategies in medical imaging, notably for brain age regression and tissue segmentation~\citep{dinsdale2021unlearning} and T1-weighted scan harmonization~\citep{cackowski_imunity_2023}. While conceptually related, these implementations differ in that the domain classifier is attached solely to the U-Net bottleneck features. Besides, \cite{dinsdale2021unlearning} injected features from the segmentation output into domain classifier, thereby unlearning also the decoder part. We experimentally found this to result in unstable training loss dynamics and worse segmentation performance, requiring extensive hyperparameter tuning to achieve a reasonable outcome.

Interestingly, \cite{dinsdale2021unlearning} reported  marginal improvements in segmentation accuracy with unlearning, and in our own experiments, applying such a strategy to WML segmentation led to performance degradation on both seen and unseen test sets (Table~\ref{table:lesion-segmentation-results-comparison}). In contrast, our proposed SSMSU approach based on multiple auxiliary classifiers enabled better control of confusion loss backpropagation and gradient loss prevention and achieved a more effective performance balance; i.e., it maintained performance on seen test data, while significantly improving all key metrics (except LTPR) on unseen cases, outperforming the five tested benchmark methods. In comparison, the SSMSU demonstrated the advantage of unlearning by enhancing lesion sensitivity (TPR, LTPR) and limiting false detections (lower LFDR), which resulted in higher overall segmentation quality in terms of DSC and RVE.

The stated hypothesis that the \textit{efficacy of strategies involving modifications or constraints on the latent space depends on the level and/or depth of segmentation model supervision during training} conjectures with our findings in Table~\ref{table:domain-accuracies-top-bot}. Model with deeper supervision achieved better domain feature disentanglement, judged by that the domain classifiers attained non-informative random output (accuracy close to 50 \%, cf. %
Figure~\ref{figure:ablation-metrics-stages-boxplots}). 
 This was not achieved when unlearning top stages (1, 2 \& 3) nor all (1--6). Furthermore, we observed larger variations in segmentation task performance when features in three top stages were exposed to domain-shift suppression. This is likely due to the stronger influence of high-resolution features on deeper layers, where suppressing domain-specific information may conflict with preserving features critical for obtaining accurate segmentation (cf. Figure~\ref{figure:ablation-metrics-stages-boxplots} with settings \textit{C} and \textit{D}). On the contrary, more subtle performance variations were observed in unlearning bottom stages (cf. Figure~\ref{figure:ablation-metrics-stages-boxplots} with settings \textit{A} and \textit{B}). 

\subsection{Comparative evaluation}

Five benchmark unsupervised domain adaptation strategies were compared: \textbf{(i)} the nnU-Net without data augmentation (no DA) establishes a baseline for domain-shift sensitivity; \textbf{(ii)} the default nnU-Net with comprehensive augmentation and test-time ensembling enhances robustness through synthetic variability but lacks explicit domain handling; \textbf{(iii)} nnU-Net with N4 bias correction and adaptive non-local means denoising reduces acquisition-related variability via  preprocessing, yet does not enforce domain-invariant representations; \textbf{(iv)} the \cite{dinsdale2021unlearning} unlearning method, adapted to the nnU-Net framework, uses adversarial training from the U-net bottleneck, but does not balanced well the main task and suppression of domain-specific features. 
This spectrum from passive to active strategies supports a comprehensive assessment of their efficacy.

On WMH's seen test subset the proposed SSMSU achieved best overall ranking, boosting TPR, LTPR and RVE. Whereas the LFDR was significantly higher than with the default nnU-Net, the DSC value was similar (0.773 vs. 0.772; no significant difference). Despite similarities, \cite{dinsdale2021unlearning} achieved worst results versus the SSMSU, which we attribute to the self-supervised scheduling of the unlearning steps that balances the trade-off between the performance of the main segmentation task and the goal of classifier confusion. Using preprocessed scans yielded poor performances, indicating that the harmonization effects due to intensity manipulation do not outweigh the loss of information relevant for segmentation task. Interestingly, the effect of preprocessing was even more detrimental compared to the nnU-Net model trained without data augmentation. 

Baseline nnU-Net without data augmentation achieved worst performance on all unseen test sets. Results improved substantially with data augmentation, consistently ranking second best (cf. Table~\ref{table:lesion-segmentation-results-comparison}). Interestingly, the same model on preprocessed images yielded inferior performance, as observed on the seen sets. Observation is consistent with previous researches that obtained good generalization of model performances without preprocessing such as skull stripping and/or bias field correction, for brain and WML segmentation \citep{cerri_contrast-adaptive_2021,wiltgen_lst-ai_2024} or other tasks like brain age regression \citep{dartora_deep_2024}.

\subsection{Annotation variability and benchmarking challenges}

Performance differences between seen and unseen sets cannot be solely attributed to the impact of domain shift. Table~\ref{table:lesion-segmentation-results-comparison} shows a substantial and consistent gap between the seen and unseen, and within unseen, subsets for each tested model or strategy. Hence, besides scanner variability, other sources of variability in publicly available datasets affect the consistency and objectivity of benchmarking efforts.

For instance, the four datasets involved different numbers of expert raters who annotated lesions using different tools and protocols, inevitably introducing inconsistencies into the ground truth or reference WML segmentations. These inconsistencies can confound performance comparisons and hinder the development of universally robust models. To enable more thorough and objective validation, there is a pressing need for larger, multi-center public datasets with standardized and consistently annotated ground truths, which would serve as reliable platforms for evaluating and advancing WML segmentation techniques.

 Compared to nnU-Net, the proposed SSMSU model achieved significantly higher DSC, TPR, LTPR, and lower RVE on the unseen test set, while exhibiting a higher LFDR. Although the increased LFDR indicates more false-positive detections, our analysis of the 51 MSSEG 2016 unseen test scans shows that nnU-Net was superior in only 5 cases, whereas SSMSU performed best in 47 cases. As illustrated in the newly added Fig.~\ref{figure:cases-6x2}, SSMSU typically produces more false-positive voxels (cf. columns 2, 3, 4, and 6), yet visual inspection reveals that many of these correspond to plausible lesions that were not annotated in the reference standard. In subjects with very few lesions, such false positives can disproportionately affect LFDR, shifting the balance in favor of the baseline. Nonetheless, the substantially lower RVE from SSMSU indicates that false positives tend to be smaller, and further post-processing (e.g.\ morphological filtering or CRF-based refinement~\citep{kamnitsas_efficient_2017}) could reduce LFDR without compromising other metrics.

 \begin{figure*}
  \centering
  \includegraphics[width=1.0\textwidth]{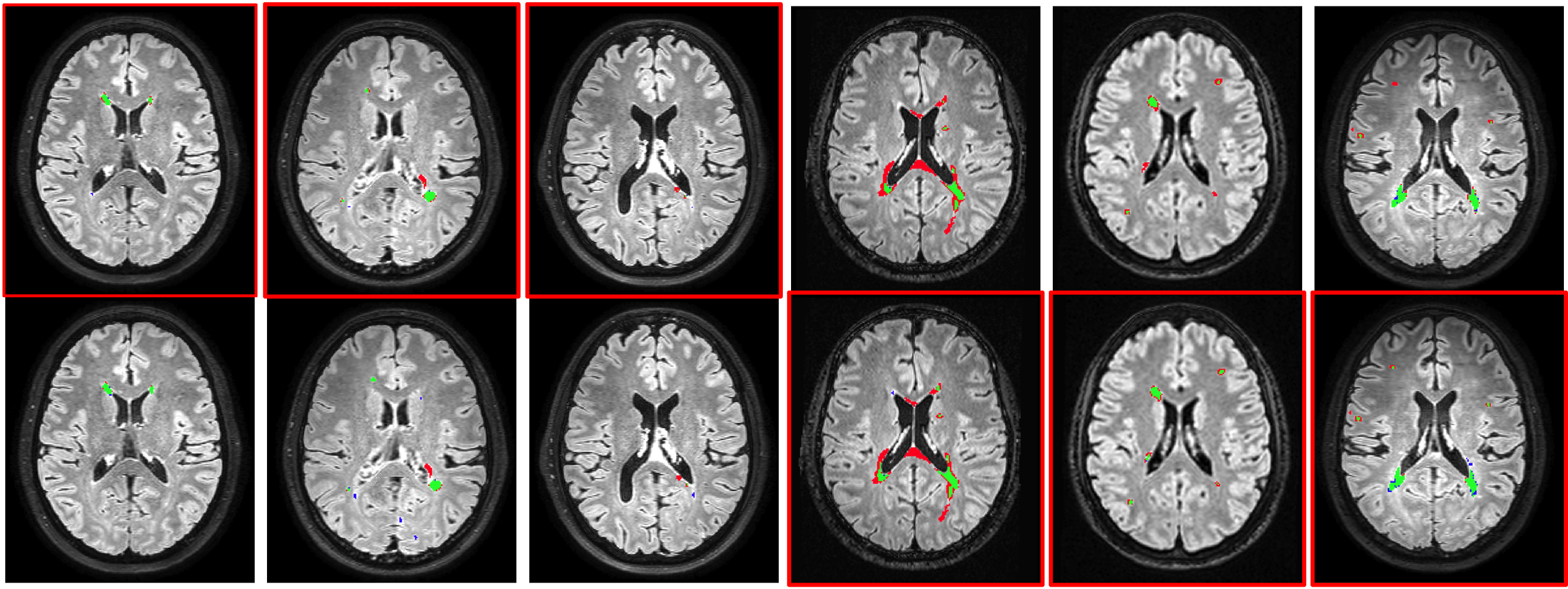}
  \caption{Axial cross-sections of six MSSEG 2016 MR scans (\textit{columns}) for the nnU-Net baseline (\textit{top row}) and the proposed nnU-Net SSMSU (\textit{bottom row}) with superimposed true positives (\textit{green}), false positives (\textit{blue}) and false negatives (\textit{red}) (\textit{bottom row}). \textit{Red frame} indicates best result according to the DSC metric. Overall, nnU-Net baseline was best in 5, while the proposed nnU-Net SSMSU was best in 47 out of 51 test cases.} \label{figure:cases-6x2}
\end{figure*}

\begin{figure}[!h]
    \centering
        \includegraphics[width=1.00\textwidth]{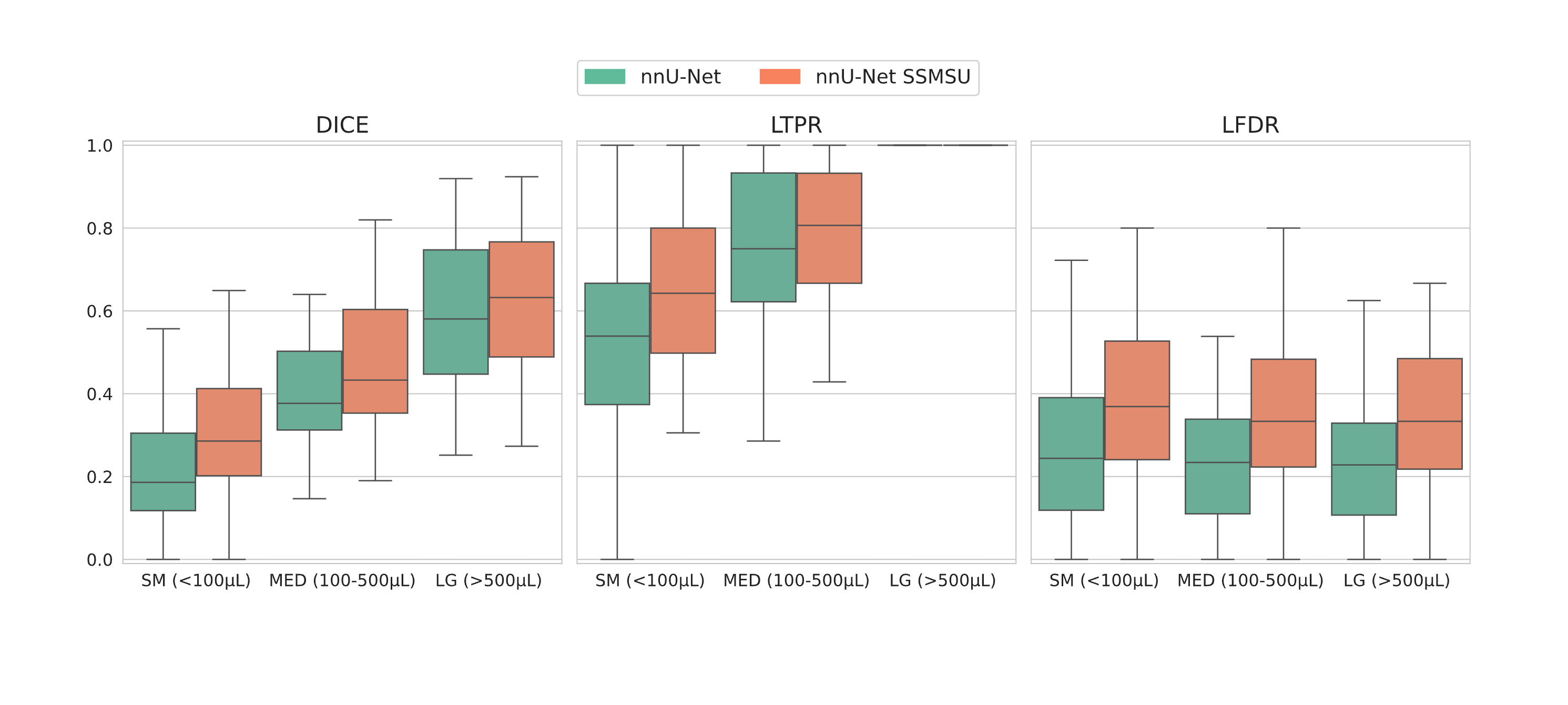}
\caption{Lesion segmentation performance according to lesion size, i.e. small (SM), medium (MED) and large (LG); obtained on the MSSEG dataset.}
    \label{fig:perf-per-lesion-size}
\end{figure}

Segmentation performance analysis by lesion size in Fig.~\ref{fig:perf-per-lesion-size} showed consistently higher DSC, LTPR and LFDR across all lesion sizes with the proposed SSMSU model, compared to nnU-Net. The former detected notably more small lesions (volume $<100 \mu{l}$) as indicated by median LTPR of $0.67$ vs. $0.55$, respectively. The LFDR value and differences were consistent across lesion size groups, i.e. median of $0.38$ vs. $0.25$ for respective SSMSU and baseline nnU-Net, likely due to ground truth limitations as illustrated in Fig.~\label{figure:cases-6x2}.

\subsection{Mitigation of study limitations}

Segmentation of WMLs is an active and competitive research field, with many effective methods available beyond those evaluated in this study. 
To enable future comparative analyses we trained and evaluated our models on four publicly available datasets (Table~\ref{table:wmh-data-description}). Furthermore, we integrated the proposed SSMSU strategy into the nnU-Net framework, which has consistently delivered state-of-the-art results across a wide range of medical segmentation challenges~\citep{isensee2024nnunetrevisitedrigorousvalidation}. The source code is publicly available\footnote{Source code is available at \url{https://github.com/Pubec/nnunetv2-unlearning}.}, enabling reproducibility, extension, and fair comparison by the broader research community.

One limitation of this study is that the proposed approach was evaluated exclusively on the task of WML segmentation. While clinically significant, this represents only one of many possible applications in neuroimaging and beyond. However, the seamless integration of SSMSU into the nnU-Net framework—along with its self-supervised training and minimal hyperparameter tuning—makes it readily applicable to other datasets and segmentation tasks. This flexibility encourages further exploration and validation of the method in broader clinical and research settings.

Another limitation is the relatively small size of the training dataset of 40 cases. This choice was intentional, as we prioritized the use of publicly available datasets to ensure reproducibility and to provide a transparent foundation that others can build upon. While this approach supports openness and comparability, it inherently restricts the biological and scanner variability available during training, which may impact segmentation performance and generalization capability. Notably, the LST-AI model~\citep{wiltgen_lst-ai_2024} was trained on a substantially larger in-house acquired and annotated dataset of 491 cases and evaluated on the same unseen datasets, reporting similar performance on MSSEG (all cases), superior on MSLJ and inferior on ISBI\footnote{Cf. Table~\ref{table:lesion-segmentation-results-comparison}, based on Supplementary Data 1 in~\citep{wiltgen_lst-ai_2024}.}. Besides larger train set, which may scale the performance, their model inputs both T1-weighted and FLAIR scans, whereas the SSMSU (and other methods in this study) input only FLAIR. Interestingly, other four methods tested in~\citep{wiltgen_lst-ai_2024} all showed inferior performances compared to SSMSU on all three unseen test sets.

These findings confirm that multimodal inputs can provide complementary information, yet they also demonstrate that a carefully designed FLAIR-only model—trained on heterogeneous public datasets and equipped with multi-stage self-supervised unlearning—can achieve performance competitive with a substantially larger multimodal system, and in some cases surpass it on unseen scanners. Moreover, the WMH dataset forming our training set contains vascular-origin lesions that differ in phenotype from MS lesions, likely increasing task difficulty on unseen MS lesions, but also limiting potential gains from additional modalities. Overall, the comparison supports our design choice: a single-modality, FLAIR-only framework offers strong cross-site robustness while avoiding the variability, missing-modality issues, and scanner-dependent uncertainty that multimodal pipelines often introduce.

Most models use a multi-modality MRI for WML segmentation, typically T1-weighted and FLAIR~\citep{wiltgen_lst-ai_2024}, or even more \citep{cerri_contrast-adaptive_2021}. We limited our experiments to using a single FLAIR modality for WML segmentation as this offers several practical and methodological advantages. FLAIR imaging provides high sensitivity and contrast for detecting WMLs, especially in periventricular and juxtacortical regions, making it well-suited as a standalone modality. Relying on one modality simplifies preprocessing pipelines, reduces inter-modality registration errors, and eliminates the need for inter-modality harmonization, which can introduce variability. Clinically, single-modality approaches increase applicability, particularly in retrospective studies or settings where only FLAIR is routinely acquired. Moreover, single-modality models are significantly less memory- and computation-intensive, enabling training and inference on a wider range of hardware, including resource-constrained environments. This increases the method’s accessibility and usability in both research and clinical settings.

While the proposed multi-stage self-supervised unlearning framework is, in principle, compatible with other forms of nuisance-variable suppression (e.g., modality, age, or sex unlearning), extending it to other anatomies or multi-modal inputs would require additional architectural considerations—such as modality-specific encoders or separate unlearning branches—which fall outside the scope of the current work. 

\subsection{Conclusion}

The proposed SSMSU segmentation method shows strong potential for clinical application in estimating lesion-based biomarkers, such as lesion volume and count, which are critical for diagnosis and monitoring treatment response~\citep{popescu_brain_2013}, and predicting MS progression~\citep{oship_assessment_2022}. SSMSU achieved the low RVE, directly reflecting good accuracy and low uncertainty in lesion volume estimation. Additionally, its performance on lesion-wise metrics -- particularly the high LTPR and balanced LFDR -- indicates reliable lesion count estimation, with high LTPR being especially important for detecting new or enlarging lesions in longitudinal studies. Taken together, our results suggest that SSMSU reasonably addresses the clinical demands for accurate and robust biomarker extraction in MS imaging.

\begin{acks}
This study was supported by the Slovenian Research Agency (Core Research Grant No. P2-0232 and Research Grant J2-3059).
\end{acks}

\bibliographystyle{ACM-Reference-Format}
\bibliography{bibliography}

@String{Computing = "Computing" }

@String{Computer = "{IEEE} Computer" }

@String{Springer = "Springer-Verlag" }

@misc{Kuijf2022_miccai2017wmh,
  doi       = {10.34894/AECRSD},
  author    = {Kuijf,  Hugo and Biesbroek,  Matthijs and De Bresser,  Jeroen and Heinen,  Rutger and Chen,  Christopher and Van Der Flier,  Wiesje and {Barkhof} and Viergever,  Max and Biessels,  Geert Jan},
  title     = {Data of the White Matter Hyperintensity (WMH) Segmentation Challenge},
  publisher = {DataverseNL},
  year      = {2022}
}

@article{Carass2017,
  title     = {Longitudinal multiple sclerosis lesion segmentation: Resource and challenge},
  volume    = {148},
  issn      = {1053-8119},
  doi       = {10.1016/j.neuroimage.2016.12.064},
  journal   = {NeuroImage},
  publisher = {Elsevier BV},
  author    = {Carass,  Aaron and Roy,  Snehashis and Jog,  Amod and Cuzzocreo,  Jennifer L. and Magrath,  Elizabeth and Gherman,  Adrian and Button,  Julia and Nguyen,  James and Prados,  Ferran and Sudre,  Carole H. and Jorge Cardoso,  Manuel and Cawley,  Niamh and Ciccarelli,  Olga and Wheeler-Kingshott,  Claudia A.M. and Ourselin,  Sébastien and Catanese,  Laurence and Deshpande,  Hrishikesh and Maurel,  Pierre and Commowick,  Olivier and Barillot,  Christian and Tomas-Fernandez,  Xavier and Warfield,  Simon K. and Vaidya,  Suthirth and Chunduru,  Abhijith and Muthuganapathy,  Ramanathan and Krishnamurthi,  Ganapathy and Jesson,  Andrew and Arbel,  Tal and Maier,  Oskar and Handels,  Heinz and Iheme,  Leonardo O. and Unay,  Devrim and Jain,  Saurabh and Sima,  Diana M. and Smeets,  Dirk and Ghafoorian,  Mohsen and Platel,  Bram and Birenbaum,  Ariel and Greenspan,  Hayit and Bazin,  Pierre-Louis and Calabresi,  Peter A. and Crainiceanu,  Ciprian M. and Ellingsen,  Lotta M. and Reich,  Daniel S. and Prince,  Jerry L. and Pham,  Dzung L.},
  year      = {2017},
  month     = mar,
  pages     = {77–102}
}

@inproceedings{wolleb2022learn2ignore,
  author    = {Wolleb, Julia
               and Sandk{\"u}hler, Robin
               and Bieder, Florentin
               and Barakovic, Muhamed
               and Hadjikhani, Nouchine
               and Papadopoulou, Athina
               and Yaldizli, {\"O}zg{\"u}r
               and Kuhle, Jens
               and Granziera, Cristina
               and Cattin, Philippe C.},
  editor    = {Wang, Linwei
               and Dou, Qi
               and Fletcher, P. Thomas
               and Speidel, Stefanie
               and Li, Shuo},
  title     = {Learn to Ignore: Domain Adaptation for Multi-site MRI Analysis},
  booktitle = {Medical Image Computing and Computer Assisted Intervention -- MICCAI 2022},
  year      = {2022},
  publisher = {Springer Nature Switzerland},
  address   = {Cham},
  pages     = {725--735}
}

@article{dinsdale2021unlearning,
  title   = {Deep learning-based unlearning of dataset bias for MRI harmonisation and confound removal},
  journal = {NeuroImage},
  volume  = {228},
  pages   = {117689},
  year    = {2021},
  doi     = {10.1016/j.neuroimage.2020.117689},
  author  = {Nicola K. Dinsdale and Mark Jenkinson and Ana I.L. Namburete}
}

@article{isensee2021nnunet,
  title     = {{nnU-Net}: a self-configuring method for deep learning-based biomedical image segmentation},
  author    = {Isensee, Fabian and Jaeger, Paul F and Kohl, Simon AA and Petersen, Jens and Maier-Hein, Klaus H},
  journal   = {Nature methods},
  volume    = {18},
  number    = {2},
  pages     = {203--211},
  year      = {2021},
  publisher = {Nature Publishing Group},
  doi       = {10.1038/s41592-020-01008-z}
}

@article{commonwick2021miccai2016,
  title   = {Multiple sclerosis lesions segmentation from multiple experts: The MICCAI 2016 challenge dataset},
  journal = {NeuroImage},
  volume  = {244},
  pages   = {118589},
  year    = {2021},
  issn    = {1053-8119},
  doi     = {10.1016/j.neuroimage.2021.118589},
  author  = {Olivier Commowick and Michaël Kain and Romain Casey and Roxana Ameli and Jean-Christophe Ferré and Anne Kerbrat and Thomas Tourdias and Frédéric Cervenansky and Sorina Camarasu-Pop and Tristan Glatard and Sandra Vukusic and Gilles Edan and Christian Barillot and Michel Dojat and Francois Cotton}
}

@article{fonov2009mniatlas,
  series  = {Organization for {Human} {Brain} {Mapping} 2009 {Annual} {Meeting}},
  title   = {Unbiased nonlinear average age-appropriate brain templates from birth to adulthood},
  volume  = {47},
  issn    = {1053-8119},
  doi     = {10.1016/S1053-8119(09)70884-5},
  journal = {NeuroImage},
  author  = {Fonov, VS and Evans, AC and McKinstry, RC and Almli, CR and Collins, DL},
  year    = {2009},
  pages   = {S102}
}

@article{guan2022survey,
  author  = {Guan, Hao and Liu, Mingxia},
  journal = {IEEE Transactions on Biomedical Engineering},
  title   = {Domain Adaptation for Medical Image Analysis: A Survey},
  year    = {2022},
  volume  = {69},
  number  = {3},
  pages   = {1173-1185},
  doi     = {10.1109/TBME.2021.3117407}
}

@article{sastregarriga-2020,
  author  = {Sastre-Garriga, Jaume and Pareto, Deborah and Battaglini, Marco and Rocca, Maria A. and Ciccarelli, Olga and Enzinger, Christian and Wuerfel, Jens and Sormani, Maria Pia and Barkhof, Frederik and Yousry, Tarek and De Stefano, Nicola and Tintoré, Mar and Filippi, Massimo and Gasperini, Claudio and Kappos, Ludwig and Río, Jordi and Frederiksen, Jette Lautrup and Palace, Jackie and Vrenken, Hugo and Montalbán, Xavier and Rovira, Alex},
  journal = {Nature Reviews Neurology},
  month   = {2},
  number  = {3},
  pages   = {171--182},
  title   = {{MAGNIMS consensus recommendations on the use of brain and spinal cord atrophy measures in clinical practice}},
  volume  = {16},
  year    = {2020},
  doi     = {10.1038/s41582-020-0314-x}
}

@article{8995481,
  author   = {Zhang, Ling and Wang, Xiaosong and Yang, Dong and Sanford, Thomas and Harmon, Stephanie and Turkbey, Baris and Wood, Bradford J. and Roth, Holger and Myronenko, Andriy and Xu, Daguang and Xu, Ziyue},
  journal  = {IEEE Transactions on Medical Imaging},
  title    = {Generalizing Deep Learning for Medical Image Segmentation to Unseen Domains via Deep Stacked Transformation},
  year     = {2020},
  volume   = {39},
  number   = {7},
  pages    = {2531-2540},
  doi      = {10.1109/TMI.2020.2973595}
}

@inproceedings{hatamizadeh2022swin,
  author    = {He, Yufan
               and Nath, Vishwesh
               and Yang, Dong
               and Tang, Yucheng
               and Myronenko, Andriy
               and Xu, Daguang},
  editor    = {Greenspan, Hayit
               and Madabhushi, Anant
               and Mousavi, Parvin
               and Salcudean, Septimiu
               and Duncan, James
               and Syeda-Mahmood, Tanveer
               and Taylor, Russell},
  title     = {SwinUNETR-V2: Stronger Swin Transformers with Stagewise Convolutions for 3D Medical Image Segmentation},
  booktitle = {Medical Image Computing and Computer Assisted Intervention -- MICCAI 2023},
  year      = {2023},
  publisher = {Springer Nature Switzerland},
  address   = {Cham},
  pages     = {416--426},
  isbn      = {978-3-031-43901-8}
}

@misc{bi2024misegnet,
  title         = {MI-SegNet: Mutual Information-Based US Segmentation for Unseen Domain Generalization},
  author        = {Yuan Bi and Zhongliang Jiang and Ricarda Clarenbach and Reza Ghotbi and Angelos Karlas and Nassir Navab},
  year          = {2024},
  eprint        = {2303.12649},
  archiveprefix = {arXiv},
  primaryclass  = {eess.IV}
}

@article{manjon_adaptive_2010,
  title   = {Adaptive non-local means denoising of {MR} images with spatially varying noise levels},
  volume  = {31},
  issn    = {1522-2586},
  doi     = {10.1002/jmri.22003},
  number  = {1},
  journal = {J Magn Reson Imaging},
  author  = {Manjón, José V. and Coupé, Pierrick and Martí-Bonmatí, Luis and Collins, D. Louis and Robles, Montserrat},
  year    = {2010},
  pages   = {192--203}
}

@article{tustison_n4itk_2010,
  title   = {{N4ITK}: improved {N3} bias correction},
  volume  = {29},
  issn    = {1558-254X},
  doi     = {10.1109/TMI.2010.2046908},
  number  = {6},
  journal = {IEEE Trans Med Imaging},
  author  = {Tustison, Nicholas J. and Avants, Brian B. and Cook, Philip A. and Zheng, Yuanjie and Egan, Alexander and Yushkevich, Paul A. and Gee, James C.},
  year    = {2010},
  pages   = {1310--1320}
}

@misc{basaran2024seghedsegmentationheterogeneousdata,
  title         = {SegHeD: Segmentation of Heterogeneous Data for Multiple Sclerosis Lesions with Anatomical Constraints},
  author        = {Berke Doga Basaran and Xinru Zhang and Paul M. Matthews and Wenjia Bai},
  year          = {2024},
  eprint        = {2410.01766},
  archiveprefix = {arXiv},
  primaryclass  = {eess.IV},
  url           = {https://arxiv.org/abs/2410.01766}
}

@inproceedings{isensee2024nnunetrevisitedrigorousvalidation,
  author    = {Isensee, Fabian
               and Wald, Tassilo
               and Ulrich, Constantin
               and Baumgartner, Michael
               and Roy, Saikat
               and Maier-Hein, Klaus
               and J{\"a}ger, Paul F.},
  editor    = {Linguraru, Marius George
               and Dou, Qi
               and Feragen, Aasa
               and Giannarou, Stamatia
               and Glocker, Ben
               and Lekadir, Karim
               and Schnabel, Julia A.},
  title     = {nnU-Net Revisited: A Call for Rigorous Validation in 3D Medical Image Segmentation},
  booktitle = {Medical Image Computing and Computer Assisted Intervention -- MICCAI 2024},
  year      = {2024},
  publisher = {Springer Nature Switzerland},
  address   = {Cham},
  pages     = {488--498},
  isbn      = {978-3-031-72114-4}
}

@article{ants,
	title = {The {ANTsX} ecosystem for quantitative biological and medical imaging},
	volume = {11},
	issn = {2045-2322},
	doi = {10.1038/s41598-021-87564-6},
	number = {1},
	journal = {Scientific Reports},
	author = {Tustison, Nicholas J. and Cook, Philip A. and Holbrook, Andrew J. and Johnson, Hans J. and Muschelli, John and Devenyi, Gabriel A. and Duda, Jeffrey T. and Das, Sandhitsu R. and Cullen, Nicholas C. and Gillen, Daniel L. and Yassa, Michael A. and Stone, James R. and Gee, James C. and Avants, Brian B.},
	month = apr,
	year = {2021},
	pages = {9068},
}

@inproceedings{confidenceintervalsuncoveredready,
  author    = {Christodoulou, Evangelia
               and Reinke, Annika
               and Houhou, Rola
               and Kalinowski, Piotr
               and Erkan, Selen
               and Sudre, Carole H.
               and Burgos, Ninon
               and Boutaj, Sofi{\`e}ne
               and Loizillon, Sophie
               and Solal, Ma{\"e}lys
               and Rieke, Nicola
               and Cheplygina, Veronika
               and Antonelli, Michela
               and Mayer, Leon D.
               and Tizabi, Minu D.
               and Cardoso, M. Jorge
               and Simpson, Amber
               and J{\"a}ger, Paul F.
               and Kopp-Schneider, Annette
               and Varoquaux, Ga{\"e}l
               and Colliot, Olivier
               and Maier-Hein, Lena},
  editor    = {Linguraru, Marius George
               and Dou, Qi
               and Feragen, Aasa
               and Giannarou, Stamatia
               and Glocker, Ben
               and Lekadir, Karim
               and Schnabel, Julia A.},
  title     = {Confidence Intervals Uncovered: Are We Ready for Real-World Medical Imaging AI?},
  booktitle = {Medical Image Computing and Computer Assisted Intervention -- MICCAI 2024},
  year      = {2024},
  publisher = {Springer Nature Switzerland},
  address   = {Cham},
  pages     = {124--132},
  isbn      = {978-3-031-72117-5}
}

@article{eshaghi_identifying_2021,
  title     = {Identifying multiple sclerosis subtypes using unsupervised machine learning and {MRI} data},
  volume    = {12},
  copyright = {2021 The Author(s)},
  issn      = {2041-1723},
  url       = {https://www.nature.com/articles/s41467-021-22265-2},
  doi       = {10.1038/s41467-021-22265-2},
  language  = {en},
  number    = {1},
  urldate   = {2024-10-20},
  journal   = {Nature Communications},
  author    = {Eshaghi, Arman and Young, Alexandra L. and Wijeratne, Peter A. and Prados, Ferran and Arnold, Douglas L. and Narayanan, Sridar and Guttmann, Charles R. G. and Barkhof, Frederik and Alexander, Daniel C. and Thompson, Alan J. and Chard, Declan and Ciccarelli, Olga},
  month     = apr,
  year      = {2021},
  note      = {Publisher: Nature Publishing Group},
  pages     = {2078}
}

@article{mcginley_diagnosis_2021,
  title      = {Diagnosis and {Treatment} of {Multiple} {Sclerosis}: {A} {Review}},
  volume     = {325},
  issn       = {0098-7484},
  shorttitle = {Diagnosis and {Treatment} of {Multiple} {Sclerosis}},
  url        = {https://doi.org/10.1001/jama.2020.26858},
  doi        = {10.1001/jama.2020.26858},
  number     = {8},
  urldate    = {2024-10-20},
  journal    = {JAMA},
  author     = {McGinley, Marisa P. and Goldschmidt, Carolyn H. and Rae-Grant, Alexander D.},
  month      = feb,
  year       = {2021},
  pages      = {765--779}
}

@article{amrita_sota_wml_seg_2020,
  author  = {Kaur, Amrita and Kaur, Lakhwinder and Singh, Ashima},
  year    = {2020},
  month   = {02},
  pages   = {1-27},
  title   = {State-of-the-Art Segmentation Techniques and Future Directions for Multiple Sclerosis Brain Lesions},
  volume  = {28},
  journal = {Archives of Computational Methods in Engineering},
  doi     = {10.1007/s11831-020-09403-7}
}

@article{commowick_objective_2018,
  title     = {Objective {Evaluation} of {Multiple} {Sclerosis} {Lesion} {Segmentation} using a {Data} {Management} and {Processing} {Infrastructure}},
  volume    = {8},
  copyright = {2018 The Author(s)},
  issn      = {2045-2322},
  url       = {https://www.nature.com/articles/s41598-018-31911-7},
  doi       = {10.1038/s41598-018-31911-7},
  language  = {en},
  number    = {1},
  urldate   = {2024-10-20},
  journal   = {Scientific Reports},
  author    = {Commowick, Olivier and Istace, Audrey and Kain, Michaël and Laurent, Baptiste and Leray, Florent and Simon, Mathieu and Pop, Sorina Camarasu and Girard, Pascal and Améli, Roxana and Ferré, Jean-Christophe and Kerbrat, Anne and Tourdias, Thomas and Cervenansky, Frédéric and Glatard, Tristan and Beaumont, Jérémy and Doyle, Senan and Forbes, Florence and Knight, Jesse and Khademi, April and Mahbod, Amirreza and Wang, Chunliang and McKinley, Richard and Wagner, Franca and Muschelli, John and Sweeney, Elizabeth and Roura, Eloy and Lladó, Xavier and Santos, Michel M. and Santos, Wellington P. and Silva-Filho, Abel G. and Tomas-Fernandez, Xavier and Urien, Hélène and Bloch, Isabelle and Valverde, Sergi and Cabezas, Mariano and Vera-Olmos, Francisco Javier and Malpica, Norberto and Guttmann, Charles and Vukusic, Sandra and Edan, Gilles and Dojat, Michel and Styner, Martin and Warfield, Simon K. and Cotton, François and Barillot, Christian},
  month     = sep,
  year      = {2018},
  note      = {Publisher: Nature Publishing Group},
  pages     = {13650}
}

@book{commowick_hal_03358968,
  title       = {MSSEG-2 challenge proceedings: Multiple sclerosis new lesions segmentation challenge using a data management and processing infrastructure},
  author      = {Commowick, Olivier and Cervenansky, Frederic and Cotton, François and Dojat, Michel},
  url         = {https://inria.hal.science/hal-03358968},
  booktitle   = {{MICCAI 2021 - 24th International Conference on Medical Image Computing and Computer Assisted Intervention}},
  address     = {Strasbourg, France},
  pages       = {126},
  year        = {2021},
  month       = Sep,
  hal_id      = {hal-03358968},
  hal_version = {v3},
  key         = {MSSEG2}
}

@article{popescu_brain_2013,
  title    = {Brain atrophy and lesion load predict long term disability in multiple sclerosis},
  volume   = {84},
  issn     = {1468-330X},
  doi      = {10.1136/jnnp-2012-304094},
  language = {eng},
  number   = {10},
  journal  = {Journal of Neurology, Neurosurgery, and Psychiatry},
  author   = {Popescu, Veronica and Agosta, Federica and Hulst, Hanneke E. and Sluimer, Ingrid C. and Knol, Dirk L. and Sormani, Maria Pia and Enzinger, Christian and Ropele, Stefan and Alonso, Julio and Sastre-Garriga, Jaume and Rovira, Alex and Montalban, Xavier and Bodini, Benedetta and Ciccarelli, Olga and Khaleeli, Zhaleh and Chard, Declan T. and Matthews, Lucy and Palace, Jaqueline and Giorgio, Antonio and De Stefano, Nicola and Eisele, Philipp and Gass, Achim and Polman, Chris H. and Uitdehaag, Bernard M. J. and Messina, Maria Jose and Comi, Giancarlo and Filippi, Massimo and Barkhof, Frederik and Vrenken, Hugo and {MAGNIMS Study Group}},
  month    = oct,
  year     = {2013},
  pmid     = {23524331},
  pages    = {1082--1091}
}

@article{alzubaidi_review_2021,
	title = {Review of deep learning: concepts, {CNN} architectures, challenges, applications, future directions},
	volume = {8},
	issn = {2196-1115},
	shorttitle = {Review of deep learning},
	doi = {10.1186/s40537-021-00444-8},
	number = {1},
	urldate = {2024-10-23},
	journal = {Journal of Big Data},
	author = {Alzubaidi, Laith and Zhang, Jinglan and Humaidi, Amjad J. and Al-Dujaili, Ayad and Duan, Ye and Al-Shamma, Omran and Santamaría, J. and Fadhel, Mohammed A. and Al-Amidie, Muthana and Farhan, Laith},
	month = mar,
	year = {2021},
	pages = {53},
}

@article{damour_underspecification_2022,
	title = {Underspecification {Presents} {Challenges} for {Credibility} in {Modern} {Machine} {Learning}},
	volume = {23},
	number = {226},
	journal = {Journal of Machine Learning Research},
	author = {D'Amour, Alexander and Heller, Katherine and Moldovan, Dan and Adlam, Ben and Alipanahi, Babak and Beutel, Alex and Chen, Christina and Deaton, Jonathan and Eisenstein, Jacob and Hoffman, Matthew D. and Hormozdiari, Farhad and Houlsby, Neil and Hou, Shaobo and Jerfel, Ghassen and Karthikesalingam, Alan and Lucic, Mario and Ma, Yian and McLean, Cory and Mincu, Diana and Mitani, Akinori and Montanari, Andrea and Nado, Zachary and Natarajan, Vivek and Nielson, Christopher and Osborne, Thomas F. and Raman, Rajiv and Ramasamy, Kim and Sayres, Rory and Schrouff, Jessica and Seneviratne, Martin and Sequeira, Shannon and Suresh, Harini and Veitch, Victor and Vladymyrov, Max and Wang, Xuezhi and Webster, Kellie and Yadlowsky, Steve and Yun, Taedong and Zhai, Xiaohua and Sculley, D.},
	year = {2022},
	pages = {1--61},
}

@article{akkus_deep_2017,
	title = {Deep {Learning} for {Brain} {MRI} {Segmentation}: {State} of the {Art} and {Future} {Directions}},
	volume = {30},
	issn = {1618-727X},
	shorttitle = {Deep {Learning} for {Brain} {MRI} {Segmentation}},
	doi = {10.1007/s10278-017-9983-4},
	language = {en},
	number = {4},
	urldate = {2024-10-23},
	journal = {Journal of Digital Imaging},
	author = {Akkus, Zeynettin and Galimzianova, Alfiia and Hoogi, Assaf and Rubin, Daniel L. and Erickson, Bradley J.},
	month = aug,
	year = {2017},
	pages = {449--459},
}

@inproceedings{milletari_v-net_2016_diceloss,
	title = {V-{Net}: {Fully} {Convolutional} {Neural} {Networks} for {Volumetric} {Medical} {Image} {Segmentation}},
	shorttitle = {V-{Net}},
	doi = {10.1109/3DV.2016.79},
	urldate = {2024-10-24},
	booktitle = {2016 {Fourth} {International} {Conference} on {3D} {Vision} ({3DV})},
	author = {Milletari, Fausto and Navab, Nassir and Ahmadi, Seyed-Ahmad},
	month = oct,
	year = {2016},
	pages = {565--571},
}

@article{Kline2005RevisitingSA_celoss,
  title={Revisiting squared-error and cross-entropy functions for training neural network classifiers},
  author={Doug M. Kline and Victor L. Berardi},
  journal={Neural Computing \& Applications},
  year={2005},
  volume={14},
  pages={310-318}
}

@article{wilcoxon_individual_1945,
	title = {Individual {Comparisons} by {Ranking} {Methods}},
	volume = {1},
	issn = {0099-4987},
	doi = {10.2307/3001968},
	number = {6},
	urldate = {2024-11-08},
	journal = {Biometrics Bulletin},
	author = {Wilcoxon, Frank},
	year = {1945},
	note = {Publisher: [International Biometric Society, Wiley]},
	pages = {80--83},
}

@article{cackowski_imunity_2023,
	title = {{ImUnity}: {A} generalizable {VAE}-{GAN} solution for multicenter {MR} image harmonization},
	volume = {88},
	issn = {1361-8415},
	shorttitle = {{ImUnity}},
	doi = {10.1016/j.media.2023.102799},
	urldate = {2025-05-13},
	journal = {Medical Image Analysis},
	author = {Cackowski, Stenzel and Barbier, Emmanuel L. and Dojat, Michel and Christen, Thomas},
	year = {2023},
	pages = {102799}
}

@article{guan_multi-site_2021,
	title = {Multi-site {MRI} harmonization via attention-guided deep domain adaptation for brain disorder identification},
	volume = {71},
	issn = {1361-8415},
	doi = {10.1016/j.media.2021.102076},
	urldate = {2025-05-13},
	journal = {Medical Image Analysis},
	author = {Guan, Hao and Liu, Yunbi and Yang, Erkun and Yap, Pew-Thian and Shen, Dinggang and Liu, Mingxia},
	year = {2021},
	pages = {102076},
}

@article{fortin_removing_2016,
	title = {Removing inter-subject technical variability in magnetic resonance imaging studies},
	volume = {132},
	issn = {1095-9572},
	doi = {10.1016/j.neuroimage.2016.02.036},
	journal = {NeuroImage},
	author = {Fortin, Jean-Philippe and Sweeney, Elizabeth M. and Muschelli, John and Crainiceanu, Ciprian M. and Shinohara, Russell T. and {Alzheimer's Disease Neuroimaging Initiative}},
	year = {2016},
	pmid = {26923370},
	pmcid = {PMC5540379},
	pages = {198--212}
}

@article{fortin_harmonization_2017,
	title = {Harmonization of multi-site diffusion tensor imaging data},
	volume = {161},
	issn = {1095-9572},
	doi = {10.1016/j.neuroimage.2017.08.047},
	journal = {NeuroImage},
	author = {Fortin, Jean-Philippe and Parker, Drew and Tunç, Birkan and Watanabe, Takanori and Elliott, Mark A. and Ruparel, Kosha and Roalf, David R. and Satterthwaite, Theodore D. and Gur, Ruben C. and Gur, Raquel E. and Schultz, Robert T. and Verma, Ragini and Shinohara, Russell T.},
	year = {2017},
	pages = {149--170}
}

@article{beer_longitudinal_2020,
	title = {Longitudinal {ComBat}: {A} method for harmonizing longitudinal multi-scanner imaging data},
	volume = {220},
	issn = {1095-9572},
	shorttitle = {Longitudinal {ComBat}},
	doi = {10.1016/j.neuroimage.2020.117129},
	journal = {NeuroImage},
	author = {Beer, Joanne C. and Tustison, Nicholas J. and Cook, Philip A. and Davatzikos, Christos and Sheline, Yvette I. and Shinohara, Russell T. and Linn, Kristin A. and {Alzheimer’s Disease Neuroimaging Initiative}},
	year = {2020},
	pages = {117129}
}

@article{pomponio_harmonization_2020,
	title = {Harmonization of large {MRI} datasets for the analysis of brain imaging patterns throughout the lifespan},
	volume = {208},
	issn = {1095-9572},
	doi = {10.1016/j.neuroimage.2019.116450},
	journal = {NeuroImage},
	author = {Pomponio, Raymond and Erus, Guray and Habes, Mohamad and Doshi, Jimit and Srinivasan, Dhivya and Mamourian, Elizabeth and Bashyam, Vishnu and Nasrallah, Ilya M. and Satterthwaite, Theodore D. and Fan, Yong and Launer, Lenore J. and Masters, Colin L. and Maruff, Paul and Zhuo, Chuanjun and Völzke, Henry and Johnson, Sterling C. and Fripp, Jurgen and Koutsouleris, Nikolaos and Wolf, Daniel H. and Gur, Raquel and Gur, Ruben and Morris, John and Albert, Marilyn S. and Grabe, Hans J. and Resnick, Susan M. and Bryan, R. Nick and Wolk, David A. and Shinohara, Russell T. and Shou, Haochang and Davatzikos, Christos},
	year = {2020},
	pages = {116450},
}

@article{kamnitsas_efficient_2017,
	title = {Efficient multi-scale {3D} {CNN} with fully connected {CRF} for accurate brain lesion segmentation},
	volume = {36},
	issn = {1361-8415},
	journal = {Medical Image Analysis},
	author = {Kamnitsas, Konstantinos and Ledig, Christian and Newcombe, Virginia F. J. and Simpson, Joanna P. and Kane, Andrew D. and Menon, David K. and Rueckert, Daniel and Glocker, Ben},
	year = {2017},
	pages = {61 -- 78},
}

@article{dartora_deep_2024,
	title = {A deep learning model for brain age prediction using minimally preprocessed {T1w} images as input},
	volume = {15},
	issn = {1663-4365},
	doi = {10.3389/fnagi.2023.1303036},
	urldate = {2025-05-14},
	journal = {Frontiers in Aging Neuroscience},
	author = {Dartora, Caroline and Marseglia, Anna and Mårtensson, Gustav and Rukh, Gull and Dang, Junhua and Muehlboeck, J.-Sebastian and Wahlund, Lars-Olof and Moreno, Rodrigo and Barroso, José and Ferreira, Daniel and Schiöth, Helgi B. and Westman, Eric and for the Alzheimer’s Disease Neuroimaging Initiative and the Australian Imaging Biomarkers {and} Lifestyle Flagship Study of Ageing and the Japanese Alzheimer’s Disease Neuroimaging Initiative and the AddNeuroMed Consortium},
	year = {2024},
	note = {Publisher: Frontiers}
}

@article{cerri_contrast-adaptive_2021,
	title = {A contrast-adaptive method for simultaneous whole-brain and lesion segmentation in multiple sclerosis},
	volume = {225},
	issn = {1053-8119},
	doi = {10.1016/j.neuroimage.2020.117471},
	urldate = {2025-05-14},
	journal = {NeuroImage},
	author = {Cerri, Stefano and Puonti, Oula and Meier, Dominik S. and Wuerfel, Jens and Mühlau, Mark and Siebner, Hartwig R. and Van Leemput, Koen},
	year = {2021},
	pages = {117471},
}

@article{oship_assessment_2022,
	title = {Assessment of {T2} lesion-based disease activity volume outcomes in predicting disease progression in multiple sclerosis over 10 years},
	volume = {67},
	issn = {2211-0348},
	doi = {10.1016/j.msard.2022.104187},
	urldate = {2025-05-15},
	journal = {Multiple Sclerosis and Related Disorders},
	author = {Oship, Devon and Jakimovski, Dejan and Bergsland, Niels and Horakova, Dana and Uher, Tomas and Vaneckova, Manuela and Havrdova, Eva and Dwyer, Michael G. and Zivadinov, Robert},
	year = {2022},
	pages = {104187}
}

@article{wiltgen_lst-ai_2024,
	title = {{LST}-{AI}: {A} deep learning ensemble for accurate {MS} lesion segmentation},
	volume = {42},
	issn = {2213-1582},
	shorttitle = {{LST}-{AI}},
	doi = {10.1016/j.nicl.2024.103611},
	urldate = {2025-05-15},
	journal = {NeuroImage: Clinical},
	author = {Wiltgen, Tun and McGinnis, Julian and Schlaeger, Sarah and Kofler, Florian and Voon, CuiCi and Berthele, Achim and Bischl, Daria and Grundl, Lioba and Will, Nikolaus and Metz, Marie and Schinz, David and Sepp, Dominik and Prucker, Philipp and Schmitz-Koep, Benita and Zimmer, Claus and Menze, Bjoern and Rueckert, Daniel and Hemmer, Bernhard and Kirschke, Jan and Mühlau, Mark and Wiestler, Benedikt},
	year = {2024},
	pages = {103611}
}

@article{lesjak_novel_2018,
	title = {A {Novel} {Public} {MR} {Image} {Dataset} of {Multiple} {Sclerosis} {Patients} {With} {Lesion} {Segmentations} {Based} on {Multi}-rater {Consensus}},
	volume = {16},
	issn = {1559-0089},
	doi = {10.1007/s12021-017-9348-7},
	number = {1},
	journal = {Neuroinformatics},
	author = {Lesjak, Žiga and Galimzianova, Alfiia and Koren, Aleš and Lukin, Matej and Pernuš, Franjo and Likar, Boštjan and Špiclin, Žiga},
	year = {2018},
	pages = {51--63},
}

@article{yang_densely_2021,
	title = {A {Densely} {Connected} {Network} {Based} on {U}-{Net} for {Medical} {Image} {Segmentation}},
	volume = {17},
	doi = {10.1145/3446618},
	number = {3},
	journal = {ACM Trans. Multimedia Comput. Commun. Appl.},
	author = {Yang, Zhenzhen and Xu, Pengfei and Yang, Yongpeng and Bao, Bing-Kun},
	month = jul,
	year = {2021},
	pages = {89:1--89:14}
}

@article{alshehri_ischemic_2024,
	title = {Ischemic {Stroke} {Segmentation} by {Transformer} and {Convolutional} {Neural} {Network} {Using} {Few}-{Shot} {Learning}},
	volume = {20},
	doi = {10.1145/3699513},
	number = {12},
	journal = {ACM Trans. Multimedia Comput. Commun. Appl.},
	author = {Alshehri, Fatima and Muhammad, Ghulam},
	month = nov,
	year = {2024},
	pages = {394:1--394:21}
}

@inproceedings{ronneberger_u-net_2015,
	address = {Cham},
	title = {U-{Net}: {Convolutional} {Networks} for {Biomedical} {Image} {Segmentation}},
	isbn = {978-3-319-24574-4},
	doi = {10.1007/978-3-319-24574-4_28},
	booktitle = {Medical {Image} {Computing} and {Computer}-{Assisted} {Intervention} – {MICCAI} 2015},
	publisher = {Springer International Publishing},
	author = {Ronneberger, Olaf and Fischer, Philipp and Brox, Thomas},
	editor = {Navab, Nassir and Hornegger, Joachim and Wells, William M. and Frangi, Alejandro F.},
	year = {2015},
	pages = {234--241},
}

@article{ganin_domain-adversarial_2016,
	title = {Domain-adversarial training of neural networks},
	volume = {17},
	issn = {1532-4435},
	number = {1},
	journal = {J. Mach. Learn. Res.},
	author = {Ganin, Yaroslav and Ustinova, Evgeniya and Ajakan, Hana and Germain, Pascal and Larochelle, Hugo and Laviolette, François and Marchand, Mario and Lempitsky, Victor},
	month = jan,
	year = {2016},
	pages = {2096--2030},
}

@misc{tzeng_deep_2014,
	title = {Deep {Domain} {Confusion}: {Maximizing} for {Domain} {Invariance}},
	shorttitle = {Deep {Domain} {Confusion}},
	url = {http://arxiv.org/abs/1412.3474},
	doi = {10.48550/arXiv.1412.3474},
	urldate = {2025-12-05},
	publisher = {arXiv},
	author = {Tzeng, Eric and Hoffman, Judy and Zhang, Ning and Saenko, Kate and Darrell, Trevor},
	month = dec,
	year = {2014},
	note = {arXiv:1412.3474 [cs]},
}

@inproceedings{tishby_deep_2015,
	title = {Deep learning and the information bottleneck principle},
	url = {https://ieeexplore.ieee.org/document/7133169},
	doi = {10.1109/ITW.2015.7133169},
	urldate = {2025-12-05},
	booktitle = {2015 {IEEE} {Information} {Theory} {Workshop} ({ITW})},
	author = {Tishby, Naftali and Zaslavsky, Noga},
	month = apr,
	year = {2015},
	pages = {1--5},
}

@article{soviany_curriculum_2022,
	title = {Curriculum {Learning}: {A} {Survey}},
	volume = {130},
	issn = {1573-1405},
	shorttitle = {Curriculum {Learning}},
	url = {https://doi.org/10.1007/s11263-022-01611-x},
	doi = {10.1007/s11263-022-01611-x},
	language = {en},
	number = {6},
	urldate = {2025-12-05},
	journal = {International Journal of Computer Vision},
	author = {Soviany, Petru and Ionescu, Radu Tudor and Rota, Paolo and Sebe, Nicu},
	month = jun,
	year = {2022},
	pages = {1526--1565},
}

@article{billot_synthseg_2023,
	title = {{SynthSeg}: {Segmentation} of brain {MRI} scans of any contrast and resolution without retraining},
	volume = {86},
	issn = {1361-8415},
	shorttitle = {{SynthSeg}},
	doi = {10.1016/j.media.2023.102789},
	journal = {Medical Image Analysis},
	author = {Billot, Benjamin and Greve, Douglas N. and Puonti, Oula and Thielscher, Axel and Van Leemput, Koen and Fischl, Bruce and Dalca, Adrian V. and Iglesias, Juan Eugenio},
	year = {2023},
	pages = {102789},
}

@misc{kuang_mscda_2023,
	title = {{MSCDA}: {Multi}-level {Semantic}-guided {Contrast} {Improves} {Unsupervised} {Domain} {Adaptation} for {Breast} {MRI} {Segmentation} in {Small} {Datasets}},
	shorttitle = {{MSCDA}},
	doi = {10.48550/arXiv.2301.02554},
	publisher = {arXiv},
	author = {Kuang, Sheng and Woodruff, Henry C. and Granzier, Renee and Nijnatten, Thiemo J. A. van and Lobbes, Marc B. I. and Smidt, Marjolein L. and Lambin, Philippe and Mehrkanoon, Siamak},
	year = {2023},
}

@article{liu_learning_2024,
	title = {Learning multi-site harmonization of magnetic resonance images without traveling human phantoms},
	volume = {3},
	copyright = {2024 The Author(s)},
	issn = {2731-3395},
	doi = {10.1038/s44172-023-00140-w},
	number = {1},
	journal = {Communications Engineering},
	author = {Liu, Siyuan and Yap, Pew-Thian},
	year = {2024},
	note = {Publisher: Nature Publishing Group},
	pages = {1--10},
}

\end{document}